\documentclass[useAMS,usenatbib]{mn2e}

\usepackage{graphicx}
\usepackage[fleqn]{amsmath}

\newcommand{\eb}{e_\mathrm{b}}
\newcommand{\qb}{q_\mathrm{b}}
\newcommand{\ab}{a_\mathrm{b}}
\newcommand{\pb}{P_\mathrm{b}}
\newcommand{\ra}{\left(\frac{r}{\ab}\right)}

\newcommand{\Eg}{E_\mathrm{g}}
\newcommand{\tauf}{\tau_\mathrm{f}}
\newcommand{\ef}{e_\mathrm{f}}
\newcommand{\ms}{\mathrm{M_\odot}}
\newcommand{\eg}{e_\mathrm{g}}
\newcommand{\mps}{m\,s$^{-1}$}

\title[Planetesimal and Gas Dynamics in Binaries]{Planetesimal and Gas Dynamics in Binaries}
\author[S.-J. Paardekooper, P. Th\'ebault and G. Mellema]{S.-J. Paardekooper$^{1}$\thanks{E-mail:
S.Paardekooper@damtp.cam.ac.uk}, P. Th\'ebault$^{2,3}$ and G. Mellema$^{2}$\\
$^{1}$DAMTP, University of Cambridge, Wilberforce Road, Cambridge CB3 0WA, United Kingdom\\
$^{2}$Stockholm Observatory, Albanova Universitetcentrum, SE-10691 Stockholm, Sweden\\
$^{3}$LESIA, Observatoire de Paris, Section de Meudon, F-92195 Meudon Principal Cedex, France}
\begin{document}

\date{Draft version \today}

\pagerange{\pageref{firstpage}--\pageref{lastpage}} \pubyear{2008}

\maketitle

\label{firstpage}

\begin{abstract}
Observations of extrasolar planets reveal that planets can be found in
close binary systems, where the semi-major axis of the binary orbit is
less than 20 AU. The existence of these planets challenges planet
formation theory because the strong gravitational perturbations due to
the companion increase encounter velocities between planetesimals and
make it difficult for them to grow through accreting collisions. We
study planetesimal encounter velocities in binary systems, where the
planetesimals are embedded in a circumprimary gas disc that is allowed
to evolve under influence of the gravitational perturbations of the
companion star. We use the RODEO method to evolve the vertically
integrated Navier-Stokes equations for the gas disc. Embedded within
this disc is a population of planetesimals of various sizes, that
evolve under influence of the gravitational forces of both stars and
friction with the gas. The equations of motion for the planetesimals
are integrated using a $4^{\mathrm{th}}$ order symplectic
algorithm. We find that the encounter velocities between planetesimals
of different size strongly depend on the gas disc
eccentricity. Depending on the amount of wave damping, we find two
possible states of the gas disc: a quiet state, where the disc
eccentricity reaches a steady state that is determined by the forcing
of the binary, for which the encounter velocities do not differ more
than a factor of 2 from the case of a circular gas disc, and an
excited state, for which the gas disc obtains a large free
eccentricity, which drives up the encounter velocities more
substantially. In both cases, inclusion of the full gas dynamics
increases the encounter velocity compared to the case of a static,
circular gas disc. Full numerical parameter exploration is still
impossible, but we derive analytical formulae to estimate encounter
velocities between bodies of different sizes given the gas disc
eccentricity. The gas dynamical evolution of a protoplanetary disc in
a binary system tends to make planetesimal accretion even more
difficult than in a static, axisymmetric gas disc.
\end{abstract}


\section{Introduction}

Planetary formation in binary systems is an issue which has gained
considerable attention in recent years, fueled by the discovery of
more than 40 extrasolar planets in multiple systems, making up over
$20\%$ of the total number of exoplanets detected so far
\citep{ragha06,desi07}. Note, however, that the vast majority of these
planets have been detected in wide binaries, with separations $\geq
100\,$AU, for which the binarity of the system is probably only weakly
felt at the location of the planets (most of which have semi-major
axis $a_\mathrm{p}<3$ AU).  Nevertheless, a handfull of planets have
been identified in ``tight'' binaries with separations of typically
$\simeq 20$ AU: HD\,41004, Gliese 86 and perhaps the most famous,
$\gamma$ Cephei. For these systems there is little doubt that the
presence of the companion must have played a crucial role.

The first studies of the planets-in-binaries issue were devoted to the
problem of long term orbital stability. In this field, the reference
work probably remains the analyical and numerical exploration by  \citet{holw99}, 
which enabled these authors to derive semi-empirical
formulae for the critical semi-major axis $a_\mathrm{crit}$ beyond
which a planetary orbit becomes unstable in a binary of mass ratio
$\mu=m_2/(m_1+m_2)$, separation $\ab$ and eccentricity $\eb$. Since
then other numerical studies further explored this issue, and the
criteria for orbital stability in binaries are now fairly well
constrained \citep[e.g.][]{dav03,fat06,mud06}
\footnote{In this respect, it is worth noticing that all planets discovered
so far in binary systems are reassuringly on stable orbits}.

The problem of the binarity's influence on planet $formation$ is
a different issue. In the now ``standard'' core-accretion scenario,
planets are believed to appear through a succession of different stages,
leading in a complex and still only partially understood way
from small grains condensed in the initial nebula to
fully formed planetary objects \citep[e.g][]{lis93}.
The way each of these stages can be affected by the presence
of a companion star requires in principle a full study in itself.
In this respect, two stages have been investigated in particular.
One of them is the final mutual assembly of protoplanetary embryos into
planets \citep[e.g.][]{cham02,quin07,hag07}. This phase is almost
entirely governed by gravitation, more exactly
by the coupling between mutual gravitational pertubartions
among the embryos and the external pull of the companion star. \citet{quin07}
showed that, as a first approximation, the final stages can proceed
unimpeded at distance $a_\mathrm{p}$ if the binary's periastron $\ab$ exceeds
$\simeq\,5\,a_\mathrm{p}$.

\subsection{Planetesimal accretion}

The other stage that has been extensively studied, and the one which
we focus on in the present study, is the earlier phase leading from
kilometre-sized planetesimals to 100-500\,km sized embryos.  This
planetesimal accretion phase should in principle be much more
sensitive to the perturbing presence of the companion star. Indeed,
``normal'' planetesimal accumulation around a single star should
proceed in a dynamically cold environment, where average mutual
encounter velocities $\langle \Delta v \rangle$ are of the order of
planetesimal surface escape velocities $v_\mathrm{esc}$, i.e.  only a
few \mps\ for km-sized bodies. In such a quiet swarm, a few isolated
seed objects, initially just a little bigger than the rest of the
swarm, can grow very quickly, through the so-called runaway growth
mode \citep[e.g][]{gre78,wet89}.  This very fast and efficient
mechanism takes place because these bodies' collisional cross section
$\sigma_\mathrm{coll}$ is amplified, with respect to their geometrical
cross section $\sigma_\mathrm{geom}$, by a gravitational focusing
factor:
\begin{equation}
\frac{\sigma_\mathrm{coll}}{\sigma_\mathrm{geom}} =\left[1\,\,\,\,\,+\,\,
\,\,\left(\frac{v_{\mathrm{esc,R}}}{\langle \Delta v \rangle}\right)^{2}\right]
\simeq \left[1\,\,\,\,\,+\,\,
\,\,\left(\frac{v_{\mathrm{esc,R}}}{v_{\mathrm{esc,r}}}\right)^{2}\right]
\label{focus}
\end{equation}
where ``R'' refers to the seed object and ``r'' to the bulk of the
planetesimal population.  This growth is self-accelerating: faster
growth leads to higher amplification factor, which in its turn leads to
faster growth, and so on.  As can be seen from Eq. \eqref{focus},
runaway growth is very sensitive to $\langle \Delta v \rangle$. In a
binary system, this growth mode is obviously at risk: if the companion
is efficient enough in stirring up the planetesimal swarm, up to
velocities exceeding $v_{\mathrm{esc}(R)}$, then runaway accretion
could easily be stopped. Further stirring could even lead to total
stop of any form of accretion, by increasing $\langle \Delta v
\rangle$ above the threshold velocity $v_\mathrm{ero}$, for which all
impacts lead to mass erosion instead of accretion.

\subsection{Planetesimal impact velocities in binaries}

This possible effect of a stellar companion on the impact velocity
distribution was explored in several previous works. The first studies
dealt with the simplified case where only the gravitational potential
of the two stars is taken into account \citep{hep78,whit98}. However,
within current planet formation scenarios, it is very likely that
planetesimal accretion takes place while vast quantities of
primordial gas are left in the system.  The complex coupled effects
between binary secular perturbations and gaseous friction on the
planetesimals were numerically explored by \citet{mascho00} and
\citet{the04} for the specific cases of the $\alpha$-Centauri and
$\gamma$~Cephei systems, and more recently by \citet{the06} for the
general case of any binary configuration.  The latter study in
particular showed that, while gas drag is able, through strong orbital
alignement, to counteract the velocity stirring effect of an external
perturber for impacts between equal-sized bodies, it tends to {\it
increase}\/ $\langle \Delta v \rangle$ between objects of different
sizes. This is because gaseous friction effects are size dependent.
The main conclusion was that even relatively wide binaries with
$10\leq \ab \leq 50$ AU could have a strong accretion inhibiting
effect on a swarm of colliding kilometre-sized planetesimals
\citep[see Fig.8 of][]{the06}.

However, because of numerical contraints, these studies were carried
out assuming a perfectly axisymetric gas disc. This simple
approximation is unlikely to be realistic. Indeed, the gas should also
``feel'' the companion star's perturbations and depart from an
axisymmetric structure with circular streamlines.  To what extent does
the gas disc respond to the secular perturbations?  Will gas
streamlines align with planetesimal forced orbits, thus strongly
reducing gas drag effects, or will they significantly depart from
forced orbits, thus playing a major role for planetesimal dynamics?
More realistic models of the planetesimal+gas evolution are definitely
needed.  A pioneering attempt at modeling coupled planetesimal/gas
systems was recently performed by \citet{pawel07}. This study was
however restricted to a circular binary, which is the most simple to
numerically tackle, because in this case the spiral wave pattern that
is excited in the gas disc is stationary, which removes the need for
following the evolution of the gas numerically. Note, however, that
possible gas eccentricity growth through the effects of the 3:1
Lindblad resonance \citep{lubow91,goodchild06} is not included using
this method, nor are any effects of the viscous overstablility
\citep{kato78,latter06}. Furthermore, \citet{pawel07} focused on
following semi-major axis and eccentricity evolutions, without
computing encounter velocities estimates, which is, as previously
discussed, the crucial parameter for investigating the
accretional evolution of the swarm.  \citet{kleynel07} performed a
full scale planetesimal+gas study in eccentric binaries, including the
gravitational pull of the gas disc on planetesimals.  However, the
complexity of their numerical code made that only a limited number of
planetesimals could be followed (less than a hundred), which is not
enough to derive relative velocity statistics. Basically, including
the gravitational pull of the gas for $\sim 10000$ particles is
equivalent to including self-gravity of the gas, which is known to be
very expensive numerically.  See Sect. \ref{secDisc} for a discussion.

\subsection{Paper outline}

We aim to extend the study of \cite{the06} by including the effects of
an evolving gas disc.  Solving the hydrodynamical equations of motion
for the gas is computationally expensive compared to solving the
equations of motion for the planetesimals, which makes it impossible
to perform a parameter study as presented in \cite{the06}.  Instead,
we focus on two binary configurations only, one corresponding to a
tight binary of separation $\ab=$10\,AU with $\eb=0.3$, and the other to the specific
case of $\gamma$~Cephei, and point out the general trends induced by
dynamical evolution of the gas.

The plan of this paper is as follows. In Sect. \ref{secModel} we
discuss our model and the numerical method. In Sect. \ref{secAna} we
present some simple analytical considerations on the problem, after
which in Sect. \ref{secTest} we test our method by reproducing some of
the results of \cite{the06}. We come to the results in
Sect. \ref{secRes}, which we subsequently discuss in
Sect. \ref{secDisc}, and we sum up our conclusions in
Sect. \ref{secCon}.


\section{Model}
\label{secModel}

Our model consists of three distinct components: the binary system,
the planetesimal disc and the gas disc. All orbits are assumed to be
coplanar.

\subsection{Particles}

The orbits of the binary stars of masses $M_\mathrm{p}$ and
$M_\mathrm{b}$ (for the primary and the secondary, respectively) and
the planetesimals are integrated using a fourth order symplectic
method. The binary stars feel only the gravitational force from each
other, while the planetesimals in addition feel friction with the gas
(see Eq.  \eqref{eqNonLinDrag}):
\begin{equation}
{\vec f_\mathrm{drag}}=-K \left| \Delta {\vec v} \right| \Delta {\vec v},
\end{equation}
where $\Delta {\vec v}$ denotes the velocity difference between the
planetesimals and the gas, and the constant of proportionality $K$ is
given by:
\begin{equation}
K=\frac{3\rho_\mathrm{g}C_\mathrm{d}}{8\rho_\mathrm{pl} R},
\end{equation}
in which $\rho_\mathrm{g}$ is the gas density, $\rho_\mathrm{pl}$ is
the planetesimal internal density and $R$ is the radius of the
planetesimal. We consider spherical bodies only, which makes the drag
coefficient $C_\mathrm{d}\approx 0.4$. Following \cite{the06}, we take
$\rho_\mathrm{pl}=3$ $\mathrm{g ~cm^{-3}}$.
 
\subsection{Gas disc}
The gas disc is modeled as a two-dimensional, viscous accretion
disc. We will work in a cylindrical coordinate frame $(r,\varphi)$
with the primary star in the origin. Note that this is not an inertial
frame. It is convenient to work with units in which
$GM_\mathrm{p}=\ab=1$. In these units, the orbital frequency of the
binary is unity.

The disc extends radially from $r=0.025\ab$ to $r=0.4\ab$.
The equation of state is taken to be locally isothermal:
\begin{equation}
p=c_\mathrm{s}^2 \Sigma,
\end{equation}
where $p$ is the (vertically integrated) pressure, $\Sigma$ is the
surface density and $c_\mathrm{s}$ is the sound speed, which is
related to the disc thickness $H$ and the Kepler frequency
$\Omega_\mathrm{K}$ through
\begin{equation}
c_\mathrm{s}=H\Omega_\mathrm{K}.
\end{equation}
In all simulations presented here, we have used $H=0.05 r$.

We do not consider self-gravity for the gas, to keep the problem
tractable computationally.  Note, however, that for globally eccentric
gas discs self-gravity may be an important effect (see
Sect. \ref{secDisc}).

The surface density follows a power law, initially, with index $-7/4$
or $-1/2$.  Because we do not include self-gravity, the density can be
scaled arbitrarily, as far as the hydrodynamical evolution is
concerned. We normalize the surface density so that
$\rho_\mathrm{g}=1.4 \cdot 10^{-9}$ $\mathrm{g ~cm^{-3}}$ at 1~AU.
This is consistent with the Minimum Mass Solar Nebula (MMSN) of
\cite{haya81}.
  
The disc is initialized to have equilibrium rotation speeds (slightly
sub-Keplerian due to the pressure gradient), with zero radial
velocity. The viscosity is modeled using the $\alpha$-prescription,
with $\alpha=0.004$ , which is a standard value for simulations of
protoplanetary discs \citep[see for example][]{kley99}.
 
\subsection{Numerical method}

We use the RODEO method for solving the hydrodynamic equations
\citep{rodeo}.  This method has been used to simulate planets embedded
in protoplanetary discs consisting of gas and dust \citep{dustflow},
and was compared to other methods in \cite{testproblem}. The inclusion
of particles into the method was discussed in \cite{dustacc}.

We monitor encounters between planetesimals using the procedure which
is standard in such studies, i.e., assigning an inflated radius
$R_\mathrm{infl}$ to every particle.  When two inflated particles
intersect, we measure their relative velocity $\Delta {\vec v}$. This
gives us a distribution in space and time of collision velocities.
For a planetesimal disc consisting of $10000$ particles,
$R_\mathrm{inf}=10^{-3}$~AU resulted in good enough collision
statistics, without introducing an artificial bias in relative
velocity estimates exceeding $\sim 10\,$\mps.

We use non-reflecting boundary conditions throughout this paper
\citep[see][]{rodeo}.  This way, we avoid spurious reflections of
spiral waves off the grid boundaries. We vary the flux limiter, which
basically controls the amount of numerical wave damping, between the
minmod limiter and the soft superbee limiter (see Appendix A and
\cite{rodeo} for their definition). We also included a kinematic
$\alpha$-viscosity, parametrized by $\alpha=0.004$. The adopted grid
size is $(n_r,n_\phi)=(192,448)$, yielding a resolution $\Delta
r=0.002$ $\ab$.


\section{Analytical considerations}
\label{secAna}

In linear theory, the response of the gas disc to a gravitational
perturbation occurs only at Lindblad and corotation resonances. At
Lindblad resonances, the perturbations take the form of a trailing
spriral waves \citep{GT79} with pattern speed $\Omega_{l,m}$. For a
perturber on a circular orbit, all modes have a pattern speed equal to
the orbital frequency of the perturber $\Omega_\mathrm{b}$. A key
result is that even for a very close binary of $a_\mathrm{b}=10$~AU
the orbital frequency in the planet-forming region (say 1~AU) is much
larger than the pattern speed of the perturbation, the effect of which
is that planetesimals at 1~AU will sweep through the spiral pattern
very quickly, and experience many short encounters with the density
waves. In a recent study, \cite{pawel07} showed that only the very
small planetesimals ($< 100$ m) are significantly affected by the
waves.

Modes of low azimuthal wavenumber are potentially much more important
in perturbing the orbits of planetesimals. In this section, we look at
the effect of a global $m=1$ perturbation in the gas disc, or, in
other words, an eccentric gas disc, on the planetesimal swarm. We do
not specify the source of the gas eccentricity, which may be the
eccentric binary companion, the viscous overstability \citep{kato78},
or an eccentric resonance \citep{lubow91}.

\subsection{No-drag limit}

\cite{goodchild06} derived the governing equation for the secular
evolution of the complex eccentricity $E$ of a gas disc subject to an
external perturbing potential on a circular orbit. Two straightforward
modifications allow us to use their result on our planetesimal swarm
in the limit of no gas drag: we let the pressure $p \rightarrow 0$ and
introduce an $m=1$ component to the perturbing potential.  The
eccentricity equation then reads:
\begin{equation}
2r\Omega\frac{\partial E}{\partial t}=-\frac{iE}{r}\frac{\partial}{\partial r}\left(r^2\frac{\partial \Phi_2}{\partial 
r}\right)+\frac{i}{r^2}\frac{\partial}{\partial r}\left(r^2\Phi'_2\right),
\label{eqEnodrag}
\end{equation}
where $\Omega$ is the angular velocity and $\Phi_2$ and $\Phi'_2$ are
the axisymmetric and the $m=1$ component of the perturbing potential,
respectively.

The perturbing potential can be expanded in terms of a Fourier cosine
series in $\phi$, and, because we are interested in secular
perturbations only, time averaged.  We only need to retain the first
two terms in the Fourier series to lowest order in $r$
\citep{augereau04}:
\begin{eqnarray}\Phi_2=-\frac{GM_\mathrm{b}r^2}{4\ab^3
\left(1-\eb^2\right)^{3/2}},\\\Phi'_2=-\frac{3\eb GM_\mathrm{b}r^3}
{8\ab^4\left(1-\eb^2\right)^{5/2}},\end{eqnarray}
where quantities with subscript ``b'' refer to the binary companion. 

With above expressions for the perturbing potential, Eq. \eqref{eqEnodrag} can 
be integrated to
\begin{multline}
E(r,t)=\frac{5}{4}\frac{\eb}{1-\eb^2}\frac{r}{\ab}\\
\left(1-\exp\left(\frac{3\pi i\qb}{2\left(1-\eb^2\right)^{3/2}}\left(\frac{r}{\ab}\right)^{3/2}\frac{t}{\pb}\right)\right),
\end{multline}
where $\qb$ is the secondary to primary mass ratio and $\pb$ is the
orbital period of the binary, and we have taken $E(r,0)=0$. This
expression is equivalent to
\begin{eqnarray}
e(r,t)=2e_\mathrm{f}(r)\left|\sin\left(\frac{U}{2}\frac{t}{\pb}\right)\right|,\\
\tan\left(\varpi\right)=-\frac{\sin\left(U\frac{t}{\pb}\right)}{1-\cos\left(U\frac{t}{\pb}\right)}
\label{periosc},
\end{eqnarray}
where
\begin{eqnarray}
e_\mathrm{f}(r) & = & \frac{5}{4}\frac{r}{\ab}\frac{\eb}{1-\eb^2},
\label{ef}\\
U & = & \frac{3\pi}{2}\frac{\qb}{\left(1-\eb^2\right)^{3/2}}\left(\frac{r}{\ab}\right)^{3/2}.
\end{eqnarray}
When we replace $r$ with $a$ this corresponds exactly to the result
given by non-linear secular perturbation theory. The eccentricity
oscillates around the forced eccentricity $e_\mathrm{f}$, with a
spatial frequency that increases with time. In the first stages of the
system's evolution, strong orbital phasing between neighbouring
objects (see Eq. \eqref{periosc}) prevents orbital crossing. However,
at some point (which depends on the radial location within the system)
the frequency becomes high enough for planetesimals of high
eccentricity to encounter planetesimals of low eccentricity, resulting
in orbital crossing and mutual encounters at very high encounter
velocities \citep[for more details, see][]{the06}.
 
 \subsection{Linear drag}
 
Including a linear drag law into the equation of motion
\begin{equation}
{\vec f_\mathrm{drag}}=-\frac{\vec \Delta v}{\tauf},
\end{equation}  
where ${\vec \Delta v}$ is the velocity difference between gas and
planetesimals and $\tauf$ is the friction time scale, leads to an
extra term in Eq. \eqref{eqEnodrag}
\begin{equation}
-\frac{2r\Omega}{\tauf}\left(E-E_\mathrm{g}\right),
\label{eqLinDrag}
\end{equation}
where $E_\mathrm{g}$ is the eccentricity of the gas. The eccentricity evolution of 
the planetesimals now consists of a damped oscillation:
\begin{multline}
E(r,t)=\frac{15\eb\qb\ra^4\Omega\tauf+16i\Eg\left(1-\eb^2\right)^{5/2}}{12\qb\ra^3\left(1-\eb^2\right)
\Omega\tauf+16i\left(1-\eb^2\right)^{5/2}} \\
\left(1-\exp\left(\left[\frac{3\pi i\qb\ra^{3/2}}{2\left(1-\eb^2\right)^{3/2}}-
\frac{\ra^{-3/2}}{2\Omega\tauf}\right]\frac{t}{\pb}\right)\right).
\end{multline}
The presence of the damping term (the second term in square brackets)
ensures that the system will evolve towards a stationary state, with
\begin{equation}
E_0(r)=\frac{15\eb\qb\ra^4\Omega\tauf+16i\Eg\left(1-\eb^2\right)^{5/2}}{12\qb\ra^3\left(1-\eb^2\right)
\Omega\tauf+16i\left(1-\eb^2\right)^{5/2}}.
\end{equation}
The important thing to notice is that $E_0$ in general depends on
$\tauf$, which means that particles of different sizes (leading to
different values of $\tauf$) will evolve towards a different
eccentricity distribution. This differential orbital phasing will lead
to very high encounter velocities between planetesimals of different
sizes. Note, however, that for the special case of $\Eg=e_\mathrm{f}$
there is no size dependence in the eccentricity evolution. In this
special case, all planetesimal orbits will be damped towards
$E=e_\mathrm{f}$, regardless of their size. For this case, we expect
small encounter velocities.

\begin{figure}
\resizebox{\hsize}{!}{\includegraphics[]{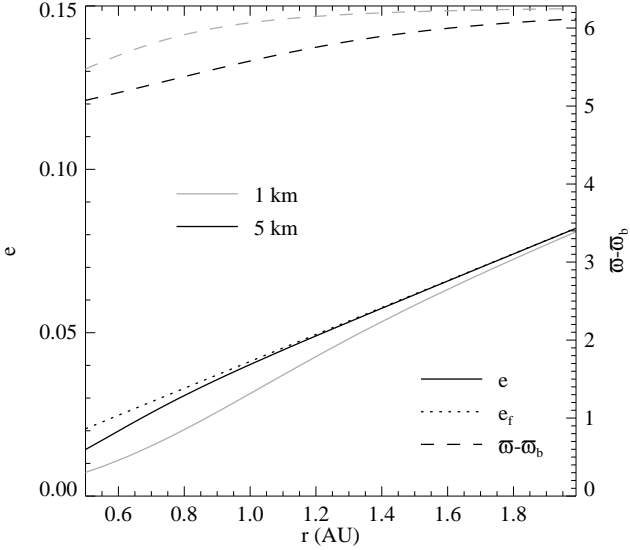}}
\caption{Axisymmetric gas disc case ($\eg=0$).  Equilibrium eccentricity
(solid curves) and $\varpi$ (dashed curves) distributions for
$\eb=0.3$, $\ab=10$~AU and $\qb=0.5$ (tight binary case) for
planetesimals of 1 km (black curves) and 5 km (gray curves), as given
by Eqs. \eqref{ecirc} and \eqref{omcirc}. Also shown is the forced
eccentricity $\ef$ (black dotted line).}
\label{figeccanapart}
\end{figure}

\subsection{Non-linear drag}

A linear drag force is not appropriate for planetesimals. For large bodies,
a non-linear drag force is more appropriate:
\begin{equation}
{\vec f_\mathrm{drag}}=-K\left|{\vec \Delta v}\right|{\vec \Delta v},
\label{eqNonLinDrag}
\end{equation} 
where
\begin{equation}
K=\frac{3\rho_\mathrm{g}}{20\rho_\mathrm{pl}R},
\end{equation}
where $\rho_\mathrm{g}$ is the gas density, $\rho_\mathrm{pl}$ is the
internal density of the planetesimals, and $R$ is the planetesimal
radius. We have assumed the planetesimals to be spherical. For this
drag law, the drag term Eq. \eqref{eqLinDrag} gets replaced by
\begin{equation}
-\sqrt{5}Kr\Omega^2\left|E-\Eg\right|\left(E-\Eg\right).
\end{equation}
In this case it is not feasible to obtain an expression for the time
evolution of the eccentricity, but we can still look at the stationary
solution, which is given by:
\begin{multline}
\frac{3\qb}{2\left(1-\eb^2\right)^{3/2}}\ra^3 E_0+
i\sqrt{5}Kr\left|E_0-\Eg\right|\left(E_0-\Eg\right)\\
-\frac{15\eb\qb}{8\left(1-\eb^2\right)^{5/2}}\ra^4=0.
\end{multline}
If we write $z=E_0-\Eg$, then
\begin{equation}
z+i\mathcal{K}\left|z\right|z+\Eg-\ef=0,
\end{equation}
with
\begin{equation}
\mathcal{K}=\frac{2\sqrt{5}Kr \left(1-\eb^2\right)^{3/2}}{3\qb \left(\frac{r}{\ab}\right)^3}.
\end{equation}
Then it is straightforward to show that in this case
\begin{equation}
E_0=\ef+\left(\Eg-\ef\right)\frac{Z-1+i\sqrt{2Z-2}}{Z+1},
\label{eqEcc}
\end{equation}
where 
\begin{equation}
Z=\sqrt{1+4\mathcal{K}^2\left| \Eg-\ef \right|^2}
\end{equation}
All size-dependence is contained in $Z$, and it is clear that again
for the special case $\Eg=e_\mathrm{f}$ there is no differential
orbital phasing. Also for this case we expect small encounter
velocities.

For completeness, we write out the magnitude $e_0$ and the argument
$\varpi_0$ of the complex equilibrium eccentricity $E_0$:
\begin{equation}
e_0^2=\frac{2\ef^2+(Z-1)\eg^2-2\sqrt{2Z-2}\ef\mathcal{I}(\Eg)}{Z+1},
\label{eqe2}
\end{equation}
\begin{multline}
\tan(\varpi_0)=-\sqrt{\frac{Z-1}{2}} \\ \frac{ (Z-1) \mathcal{I}(\Eg)
- \sqrt{2Z-2} ( \ef - \mathcal{R}(\Eg) ) } {(Z-1) \mathcal{I}(\Eg) -
\sqrt{2Z-2} \left( \ef + \mathcal{R}(\Eg) \frac{Z-1}{2}
\right)}\label{eqtanpom},
\end{multline}
where $\mathcal{R}(\Eg)$ and $\mathcal{I}(\Eg)$ denote respectively
the real and imaginary part of $\Eg$ and $\eg$ its absolute value. The
last two expressions lead to two interesting observations. When the
gas disc eccentricity vector has a significant positive imaginary
part, the eccentricity of the particles is reduced. This happens when
the gas disc is not aligned (or anti-aligned) with the binary
orbit. For the special case $Z=3$ and $\Eg=i\ef$ we have
$E_0=0$. Though this may seem a pathological case, we will see that
this behavior actually shows up in the numerical simulations.

The case of a circular gas disc (defined as having $\eg=0$, but not
necessarily being axi-symmetric) appears to be particularly simple:
\begin{eqnarray}
e_0=\sqrt{\frac{2}{1+Z}}\ef \label{ecirc} \\
\tan(\varpi_0)=-\sqrt{\frac{Z-1}{2}}. \label{omcirc}
\end{eqnarray}
In the inner regions of the disc, where gas
drag is very strong and therefore $Z \gg 1$, the eccentricity will be
much smaller than $\ef$, while the longitude of periastron will make a
90-degree angle with that of the binary. This effect can be clearly
seen in Fig. 6 of \cite{the06} (also see our Fig. \ref{figaxigas}). As $Z
\rightarrow 1$, indicating less friction (either due to a lower gas
density or a larger particle size), particles tend to adopt the forced
eccentricity and align with the binary. Note, however, that this
analysis does not take into account the oscillatory behavior of $E$,
and is therefore only valid for $Z \gg 1$ or at sufficiently late
times. We show the resulting eccentricity distribution for $\qb=0.5$,
$\eb=0.3$ and $\ab=10$~AU in Fig. \ref{figeccanapart}. These results
compare very well to the numerical results (see Fig. \ref{figaxigas}).

In the limit $\ef \ll \sqrt{Z-1} \eg$, some simple relations can be
obtained:
\begin{eqnarray}
e_0=\sqrt{\frac{Z-1}{Z+1}}\eg \label{eqee} \\
\tan(\varpi_0)=\frac{\tan(\varpi_\mathrm{g})+\sqrt{\frac{2}{Z-1}}}{1-\sqrt{\frac{2}{Z-1}}\tan(\varpi_\mathrm
{g})}, \label{eqtanpomg}
\end{eqnarray}   
where $\varpi_\mathrm{g}$ denotes the longitude of periastron of the
gas disc. We see that in the limit of strong coupling to the gas ($Z
\gg 1$), the planetesimals will tend to obtain the gas disc
eccentricity. On the contrary, for very weak coupling ($Z\rightarrow
1$), their eccentricity will tend to zero. Note, however, that
Eq. \eqref{eqee} is not valid in the limit $Z\rightarrow 1$; from
Eq. \eqref{eqe2} it is in fact easy to see that $e_0 \rightarrow \ef$
for $Z\rightarrow 1$.

An important conclusion is that, except in the special case
$\Eg=e_\mathrm{f}$, friction with an eccentric gas disc always leads
to some degree of differential orbital phasing between planetesimals
of different sizes. This case is thus qualitatively not different from
that of a circular gas disc. The magnitude of the differential orbital
phasing effect will depend strongly on the gas eccentricity though.
 
\begin{figure}
\resizebox{\hsize}{!}{\includegraphics[]{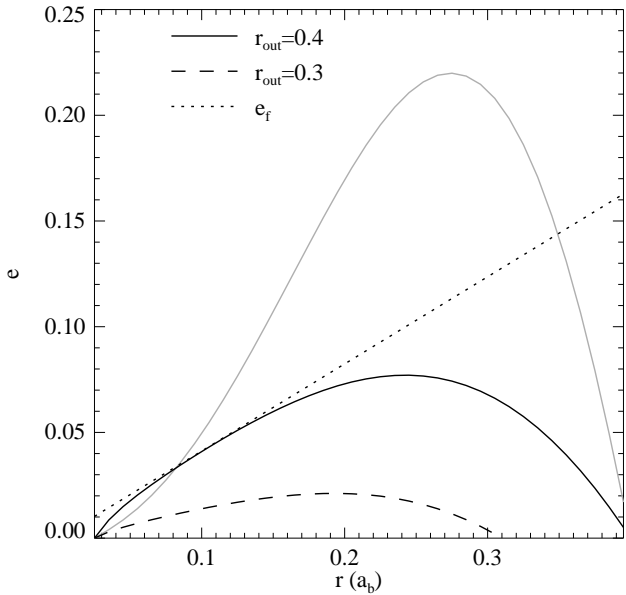}}
\caption{Equilibrium gas eccentricity distribution, analytically
derived from Eq. \eqref{eqEgas}, for $\eb=0.3$, and $\qb=0.5$ with
$\beta=-7/4$ (gray curve) or $\qb=0.234$ with $\beta=-1/2$ (black
curves). Also shown is the forced eccentricity $\ef$ (black dotted
line), which is the same for both cases. The dashed black curve shows
the effect of moving the outer boundary inwards. All models use
$h=0.05$.}
\label{figeccana}
\end{figure}

\subsection{Gas disc eccentricity}

We now consider the crucial issue of the response of the gas disc
itself to the perturbations.
The equivalent of Eq. \eqref{eqEnodrag} for a gaseous disc reads:
\begin{multline}
\label{eqEgas}
-\frac{2i}{h^2\Omega}\frac{\partial \Eg}{\partial t}=r^2\Eg''+(2+\beta)r\Eg'+\\
\left(\beta-1+\frac{3\qb}{2h^2\left(1-\eb^2\right)^{3/2}}\ra^3\right)\Eg \\
-\frac{15\eb\qb}{8h^2\left(1-\eb^2\right)^{5/2}}\ra^4,
\end{multline}
where $\beta\equiv d\log\Sigma/d\log r$, $h=H/r$, and primes denote
differentiation with respect to $r$. It is possible to find
equilibrium solutions by setting the left-hand side to zero, after
which the resulting second order ordinary differential equation can be
solved in terms of Bessel functions (for constant $\beta$). Before
doing so, we note that for the case $\beta=-1/2$, $\Eg=\ef$ is
actually a solution (for appropriate boundary conditions).

It is not straightforward to supply realistic boundary conditions for
this problem. It is conceivable that the gas disc eccentricity will
tend to zero close to the star, and numerical hydrodynamical
simulations (to which we intend to compare these results) will also
force the eccentricity to be zero at the boundaries. Therefore we
choose $\Eg(r_\mathrm{in})=0$, where $r_\mathrm{in}$ is the inner
radius of the disc. The outer boundary poses a bigger problem, because
the disc will be truncated. Since linear theory does not include the
truncation of the disc, we vary the outer boundary condition between
$\Eg(r_\mathrm{out})=0$, where $r_\mathrm{out}$ is the outer radius of
the disc and $\Eg(r_\mathrm{trunc})=0$, where $r_\mathrm{trunc}$ is
the approximate truncation radius of the disc, obtained from
hydrodynamical simulations. Note that since
$\varpi(r_\mathrm{in})=\varpi(r_\mathrm{out})=0$, the disc will be
aligned with the binary everywhere according to Eq. \eqref{eqEgas}. A
possible twist in the disc does not change the results significantly.

\begin{figure}
\resizebox{\hsize}{!}{\includegraphics[]{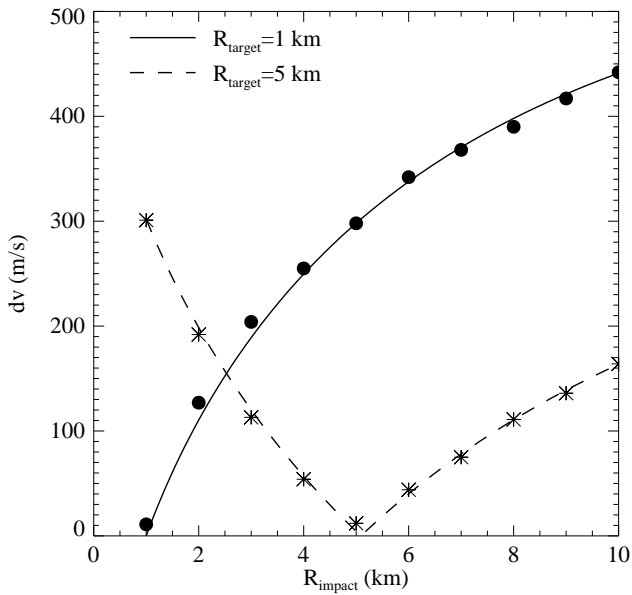}}
\caption{Circular gas disc case, for a tight binary $\qb=0.5$,
$\eb=0.3$ and $\ab=10$~AU. Shown are analytical estimates of mean
encounter velocities between planetesimals, at $1$~AU from the
primary, as a function of their size, derived using
Eq. \eqref{eqColl}.  We consider target bodies of two different sizes,
one with $R=1\,$km and the other $R=5\,$km, and derive for each case
collision velocities due to impacting objects in the $1$-$10$ km size
range.  Overplotted are the results of \citealt{the06} (their Table
$3$).  }
\label{figColl}
\end{figure}

The results are shown in Fig. \ref{figeccana}, for two binary
configurations: $\qb=0.5$, $\eb=0.3$, $\beta=-7/4$ (gray curve), and
$\qb=0.234$, $\eb=0.3$, $\beta=-1/2$ (black curves). Both cases will
be studied numerically in Sect. \ref{secRes}. We put the boundaries at
$r_\mathrm{in}=0.025$ $\ab$ and $r_\mathrm{out}=0.4$ $\ab$, the same
as will be used in Sect. \ref{secRes} in the hydrodynamical
simulations. For these binary parameters, the outer boundary is well
beyond the truncation radius, while the inner boundary is well inside
the planet-forming region (for $\ab \sim 10$~AU).

The steep density profile (denoted by the gray curve in
Fig. \ref{figeccana}) gives rise to an eccentricity well above $\ef$
for a large part of the disc. Reducing the inner radius of the disc by
a factor of 2 has virtually no effect on the eccentricity. This is not
true for the outer boundary, as we will see below. For $\beta=-1/2$,
the eccentricity approaches $\ef$ far from the boundaries. This may be
important, because in this case there would be no differential orbital
phasing in the planet-forming region. However, this strongly depends
on the location of the outer boundary. If we suppose that the disc is
circular at the truncation radius, which, for this binary
configuration, is located around $r_\mathrm{trunc}=0.3$ $\ab$, the
picture changes drastically (see the dashed line in
Fig. \ref{figeccana}). In this case, the disc remains nearly circular
everywhere, and we will see in Sect. \ref{secRes} that this closely
resembles the solution obtained from numerical hydrodynam ical
simulations. Also, there is a strong dependence on $h$, with larger
eccentricities for thinner discs.

However, there are three possibly important physical effects not
included in Eq. \eqref{eqEgas}. First of all, tidal effects are
neglected, which means that the effect of disc truncation is not taken
into account. Therefore $\beta=\beta(r)$ is unknown, and will change
dramatically near the disc edge.

There are also two possible eccentricity excitation mechanisms not
included in Eq. \eqref{eqEgas}. There is the effect of the 3:1
resonance \citep{lubow91}, and also the viscous overstability
\citep{kato78,latter06}. Both mechanisms can add a significant free
eccentricity to the disc. We rely on numerical hydrodynamical
simulations to solve for the eccentricity of the gas disc.

\begin{figure}
\resizebox{\hsize}{!}{\includegraphics[]{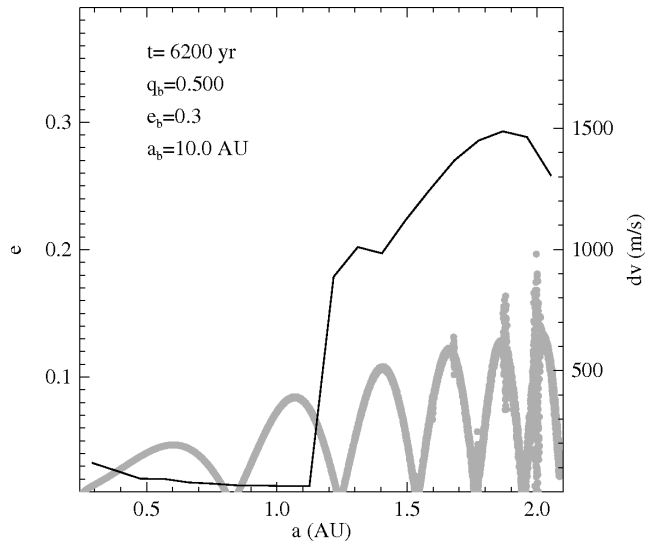}}
\caption{Planetesimal eccentricity (grey dots) and average encounter velocity
(solid line) distribution in a gas-free disc, for the tight binary case.}
\label{figgasfree}
\end{figure}

\begin{figure*}
\centering
\includegraphics[width=17cm,clip=true]{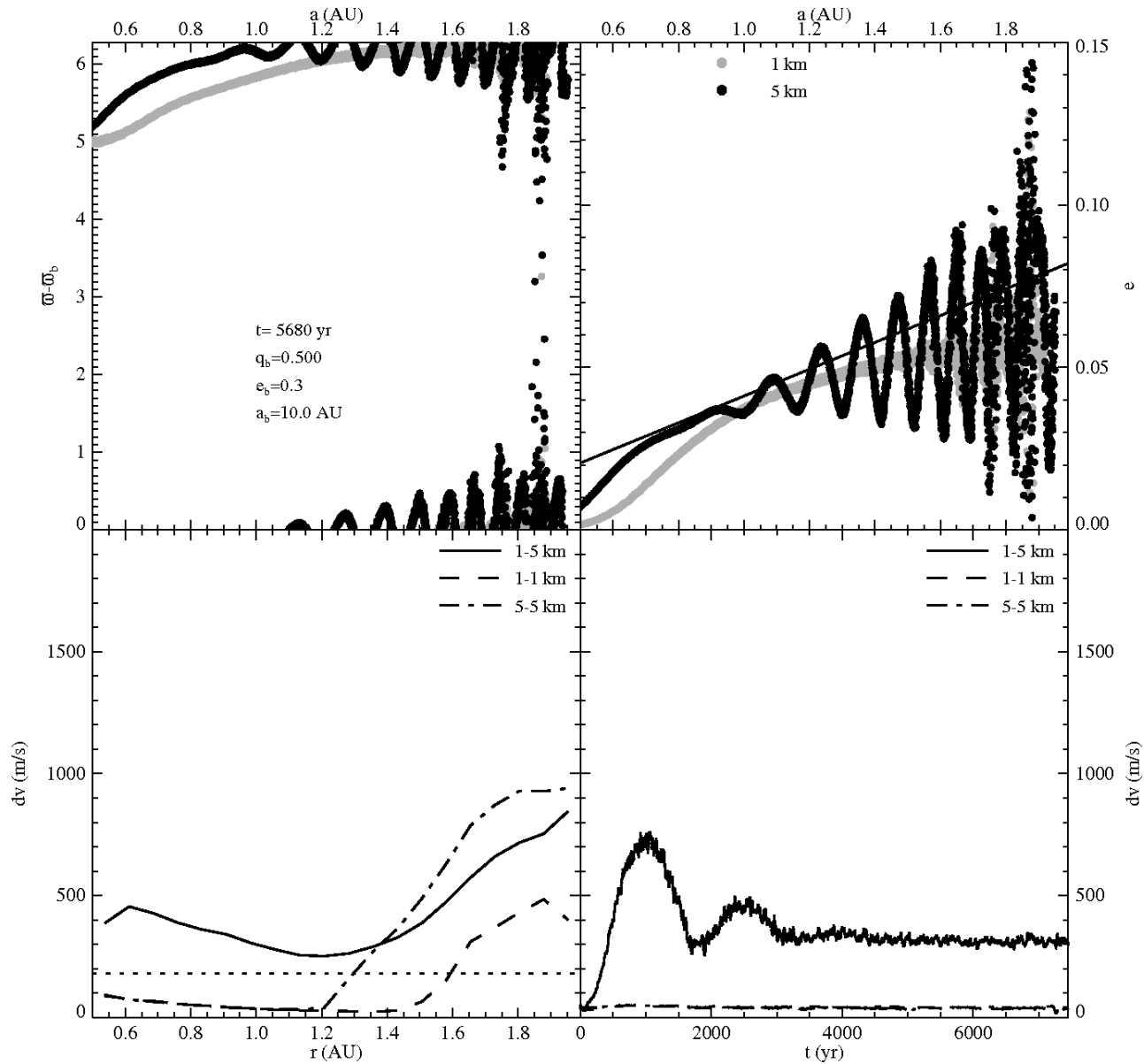}
\caption{Planetesimal evolution in an axisymmetric gas disc, with
$\Sigma_0 \propto r^{-7/4}$ for the same binary parameters as in
Fig. \ref{figgasfree} (see also the top left panel). Top left:
longitude of periastron distribution after 5680 yr. Top right:
eccentricity distribution after 5680 yr, with the black line
indicating the forced eccentricity $\ef$. Bottom left: distribution of
encounter velocities for 3 different impactor-target planetesimal
pairs: 2 corresponding to equal sized impacting objects and the third
one to a 1 km impactor hitting a 5 km target. Also shown is the
limiting maximum $\langle dv \rangle$ for accreting encounters between
1 and 5 km bodies (dotted line). Bottom right: encounter velocities at
1~AU as a function of time for the same impactor-target pairs.}
\label{figaxigas}
\end{figure*}

\subsection{Encounter velocities}

A simple model, where one considers the collision between two kinds of
bodies with the same semi-major axis in the guiding centre
approximation, suggests that their mean encounter velocity is
proportional to $|E_1-E_2|$. If we take
\begin{equation}
\overline{|\Delta {\vec v}|}=a\Omega\sqrt{\frac{5}{8}} |E_1-E_2|,
\label{eqColl}
\end{equation}
then this prediction reduces to the standard relation for bodies with
mean eccentricity $e_\mathrm{f}$ and randomized periastra of
\cite{lisste93}:
\begin{equation}
\overline{ |\Delta {\vec v}|}=a\Omega\sqrt{\frac{5}{4}} e_\mathrm{f}.
\end{equation}
When assuming a given eccentricity $\Eg$ for the gas disc,
Eq. \eqref{eqColl}, together with Eq. \eqref{eqEcc}, can be used to
predict the stationary solution for mutual encounter velocities.  For
the specific case of a circular gas disc, it gives a remarkably good
fit to the numerical results obtained, once the stationary state is
reached, by \cite{the06} (Fig. \ref{figColl}).  Note however that this
set of equations cannot predict if, and how fast the stationary
state will be reached. It also cannot account for the possibly crucial
effect of encounters due to high-eccentricity objects coming from
other regions of the system (in case of orbital crossing).  These
equations are nevertheless very useful in providing a reference to
which to compare numerical results obtained with an evolving gas disc.
Such a comparison can provide information on whether the gas
disc eccentricity dominates the collisional evolution of the
planetesimal population.

\begin{figure*}
\centering
\includegraphics[width=17cm,clip=true]{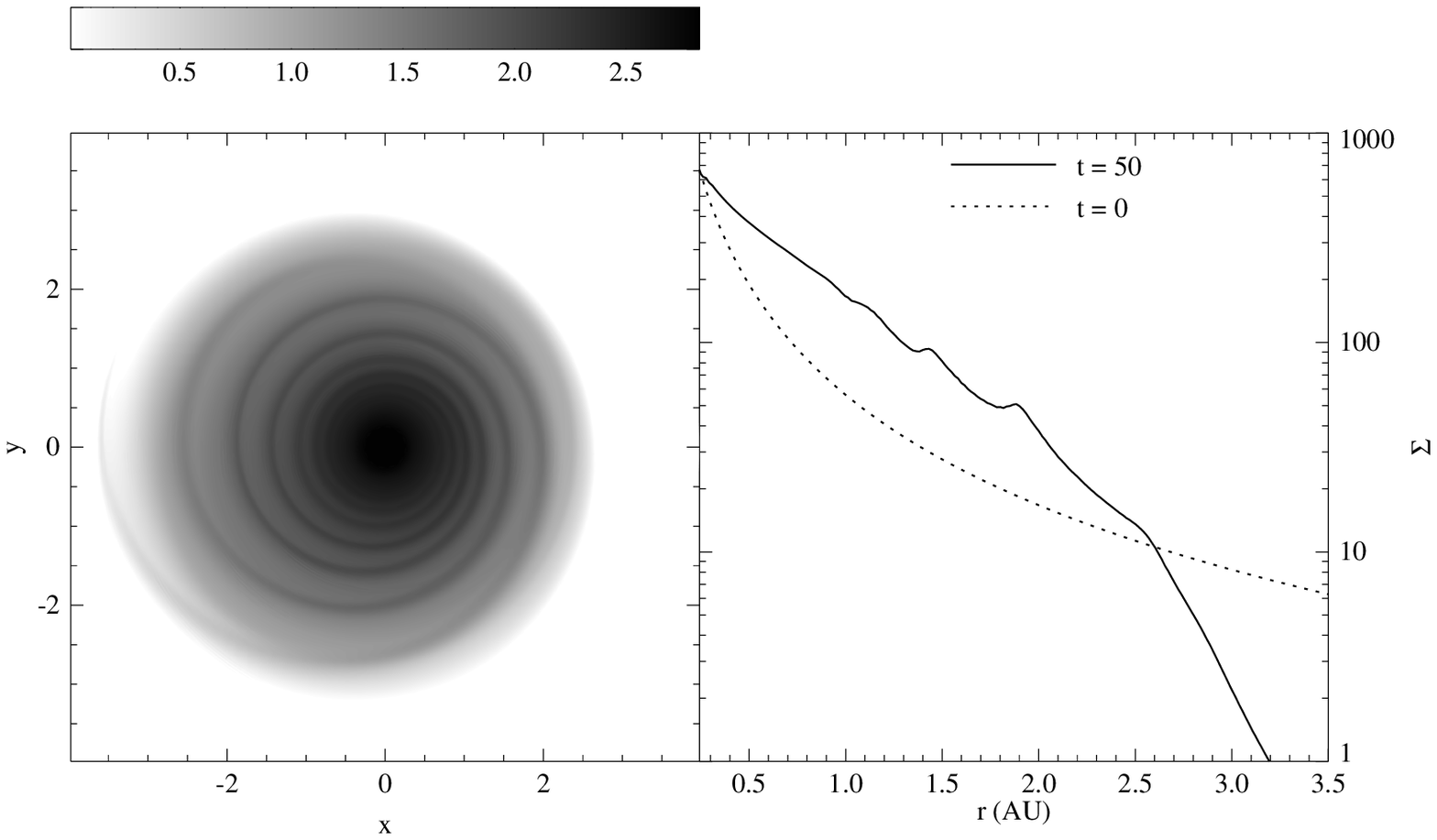}
\caption{Surface density of the gas disc, with $\Sigma_0 \propto
r^{-7/4}$ after 50 binary orbits ($\approx 1600$ yr), for the same
binary parameters as in Fig. \ref{figgasfree}. Left panel:
two-dimensional distribution of $\log_{10}\Sigma$. The binary
companion is at apo-astron, at $(x,y)=(-13,0)$. Right panel:
azimuthally averaged surface density. The density scale is arbitrary,
for comparison the initial condition is shown.}
\label{figgascont}
\end{figure*}

\section{Test cases: gas free and axisymmetric cases}
\label{secTest}

We start by reproducing the results of \cite{the06} for the gas-free
case and for an axisymmetric gas disc. These results can then serve as
a reference for the full model, but they also provide good test cases
for our method. \footnote{For the sake of comparison, we display here
results for the same specific binary configuration considered as an
example by \citet{the06}, i.e., $\ab=10$~AU, $\qb=0.5$ and
$\eb=0.3$. This corresponds to a close and eccentric binary, for which
the inner $1-2$~AU region is highly perturbed by the companion star}

\begin{figure}
\resizebox{\hsize}{!}{\includegraphics[clip=true]{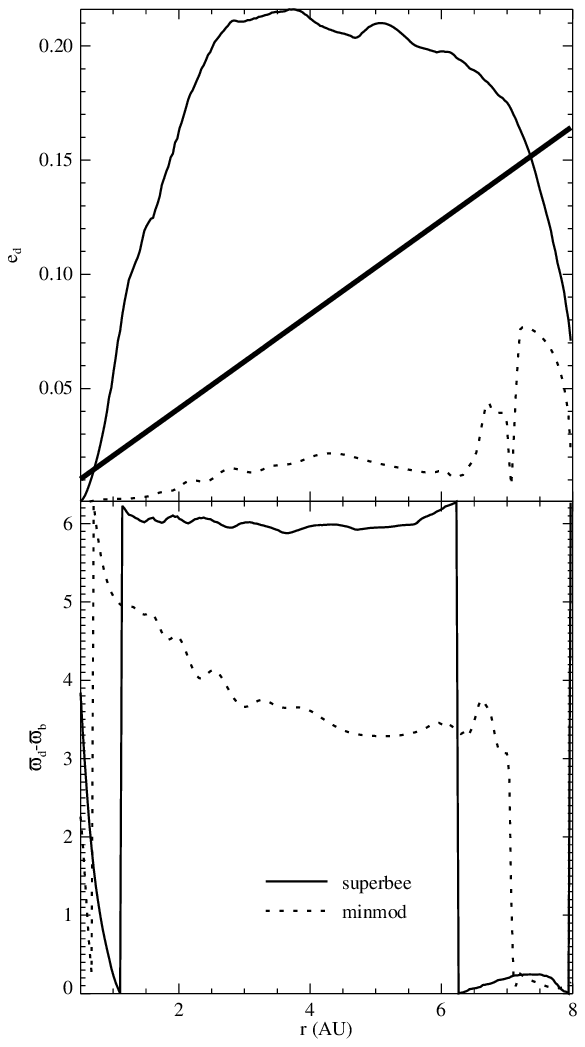}}
\caption{Gas disc eccentricity (top panel) and longitude of periastron
(bottom panel) for the excited state, obtained with the soft superbee
flux limiter (solid line) and the quiet state, obtained with the
minmod flux limiter (dotted line). The binary parameters correspond
here to the wider binary case of $\gamma$~Cephei: $\qb=0.234$,
$\ab=20$~AU, $\eb=0.3$, and the initial surface density has a shallow
radial dependence $\Sigma_0 \propto r^{-1/2}$.  The thick solid line
indicates the forced eccentricity derived from Eq. \eqref{ef}.}
\label{figgasecc}
\end{figure}

\begin{figure}
\resizebox{\hsize}{!}{\includegraphics[clip=true]{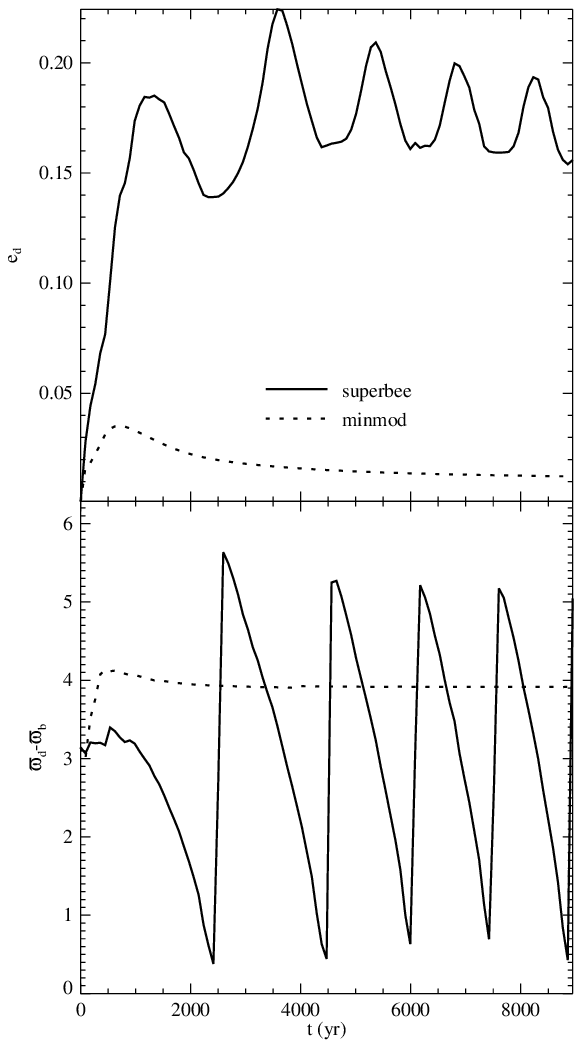}}
\caption{Evolution of the mean disc eccentricity and longitude of
periatron, for the wide binary case $\qb=0.234$, $\eb=0.3$ and
$\ab=20$~AU, with $\Sigma_0 \propto r^{-1/2}$, using two different
flux limiters.}
\label{figprecess}
\end{figure}

\subsection{Gas-free disc}

In the case of no gas drag, the eccentricity evolution of the
planetesimals is governed by secular perturbations due to the binary
only. These perturbations give rise to eccentricity oscillations, with
a wavelength that decreases with time. Due to the strong orbital
phasing large eccentricities may arise with relatively low encounter
velocities.  However, as time goes by the eccentricity oscillations
become so narrow that orbital crossing between high and low
eccetricity bodies eventually occur, resulting in a sudden and
dramatic increase of encounter velocities. Orbital crossing first
occurs in the outer regions (closer to the companion star), the radial
location at which orbital crossing occurs progressively moving inward
with time. The time at which orbital crossing occurs at a given
location can be calculated analytically following the empirical
derivation of \citet{the06}.
 
In Fig. \ref{figgasfree} we show the eccentricity distribution of
planetesimals after 6200 yr (200 binary orbits) in a tight binary
system with parameters as indicated above the figure.  The
eccentricity oscillation can be identified easily. In the outer part
of the disc, several particles have been captured into resonances. Note that
we only consider here the inner $\leq 2$~AU part of the disc, since
most planetesimal orbits beyond this distance are dynamically
unstable.

The black solid line in Fig. \ref{figgasfree} indicates the mean
encounter velocities as a function of distance. The radius where
orbital crossing occurs can be found near $1.1$~AU (note the very
sharp transition from $\langle \Delta v \rangle \simeq 0$ to very high
velocity values in less than $0.1$~AU around the crossing location.
These results quantitatively agree with the
calculations of \cite{the06} (see their Fig. 2).

\subsection{Axisymmetric gas disc}

\cite{the06} included effects of gas drag from an axisymmetric gas
disc.  In this section, we use the same gas disc model as \cite{the06}
and do not let it evolve. We divide the 10000 planetesimals into two
groups of 1 km and 5 km in size.  We consider the same tight binary
configuration as in the gas free case: $\ab=10\,$, $\eb=0.3$ and
$\qb=0.5$. In Fig. \ref{figaxigas}, we show the resulting orbital and
collisional evolution, in the $\leq 2$~AU region corresponding to
stable orbits.  These can be compared to Figs. 6 and 7 of
\cite{the06}.

In the top panels of Fig. \ref{figaxigas}, we observe the well known
\citep[e.g.][]{mascho00,the04} double effect of gas drag on planetesimal
orbits, i.e., damping of the eccentricities and periastron alignment.
At first sight, this may seem advantageous for planetesimal accretion.
However, the equilibrium between eccentricity excitation and damping
depends on the size of the planetesimals. Not only are the
eccentricities of the smaller bodies damped more readily than for the
larger bodies, they are damped towards a different equilibrium. This
is also true for the equilibrium value of the periastron alignment.
This differential phasing is responsible for high collision velocities
between particles of different sizes \citep{the06}.  Note also that in
the inner disc, where the eccentricity oscillations are damped by gas
friction, the eccentricities and longitudes of periastron of the
planetesimals agree with the analytical results in
Fig. \ref{figeccanapart} .

These consequences of gaseous friction are clearly illustrated in the
bottom panels of Fig. \ref{figaxigas} which shows the evolution of the
average encounter velocities $\langle \Delta v \rangle_{R1,R2}$, for
the cases $R_1=R_2=1$ km, $R_1=R_2=5$ km, $R_1=1$ km and $R_2=5$ km.
A first obvious effect, clearly seen for the case of equal-sized
objects, is that gas drag works against orbital crossing: stronger gas
friction moves the radius at which orbital crossing occurs outward. At
1~AU, no orbital crossing occurs after 3000 yr for both planetesimal
sizes and relative velocities are still very low (see the bottom left
panel of Fig. \ref{figaxigas}).  In sharp contrast, encounter
velocities between bodies of different size are high throughout the
disc.  In the bottom right panel of Fig. \ref{figaxigas}, we see that
the encounter velocities reach a stationary value after approximately
5000 years. The equilibrium is in perfect agreement with the result from
\cite{the06}. In the bottom left panel of Fig. \ref{figaxigas} we also show the limiting value of $\langle dv \rangle$ for accreting collisions between planetesimals of 1 and 5 km (dotted line). This is a conservative estimate, based on the most optimistic value for accretion considered by \citet{the06}.

\section{Results, full model}
\label{secRes}

\subsection{Gas disc evolution}
We start the discussion of the full model by considering the gas
evolution alone.  In Fig. \ref{figgascont} we show the surface density
of the gas disc after 50 binary orbits using the minmod flux
limiter. The disc is truncated by the binary at approximately the
radius predicted by the empirical formula of \cite{holw99}.  A large
part (approximately $20 \%$) of the material is expelled from the
system, but from the right panel of Fig. \ref{figgascont} it is also
clear that the disc is compressed: the density at 1~AU is increased by
approximately a factor of 3. This means that in order to compare with
the results of the previous section, we should rescale the density in
the full model to obtain the same magnitude of the drag force at 1~AU.

The second thing that is apparent from Fig. \ref{figgascont} is the
appearance of spiral density waves. The morphology is strongly
dependent on the phase of the binary \citep[see][]{kleynel07}, with
strong shocks appearing just after periastron.  The effect of these
waves in a circular binary on the orbital elements of planetesimals
was recently studied in \cite{pawel07}, where it was shown that only
the smallest bodies ($R\approx 100$ m) are affected.

Finally, it is clear from the left panel of Fig. \ref{figgascont} that
the gas disc becomes eccentric. This is always true, but tests
show that the two considered flux limiters give very different
results. The amplitude of the eccentricity increase depends
dramatically on the amount of wave damping, as is illustrated in
Fig. \ref{figgasecc}, where we compare the eccentricity distribution
after 50 binary orbits for the $\gamma$~Cephei binary configuration,
($\qb=0.234$, $\ab=20$~AU, $\eb=0.3$) for results obtained with our
two different flux limiters.

The diffusive minmod limiter gives rise to a low eccentricity ($\eg
\leq 0.05$ and even $\leq 0.02$ in $r<4$~AU region), stationary disc
state that we henceforth call the quiet state. It is a robust state,
numerically, to changes in resolution and boundary conditions
(reflecting, non-reflecting). In fact, this solution approximately
obeys the time-independent version of Eq. \eqref{eqEgas}, as we will
show below. The maximum eccentricity does depend on the location of
the inner boundary, however, since the boundary conditions enforce the
eccentricity to be zero at the boundary. We have found that the
location of the inner boundary should be at least as far in as $0.025$
$\ab$ to get converged results.

\begin{figure*}
\centering
\includegraphics[width=17cm,clip=true]{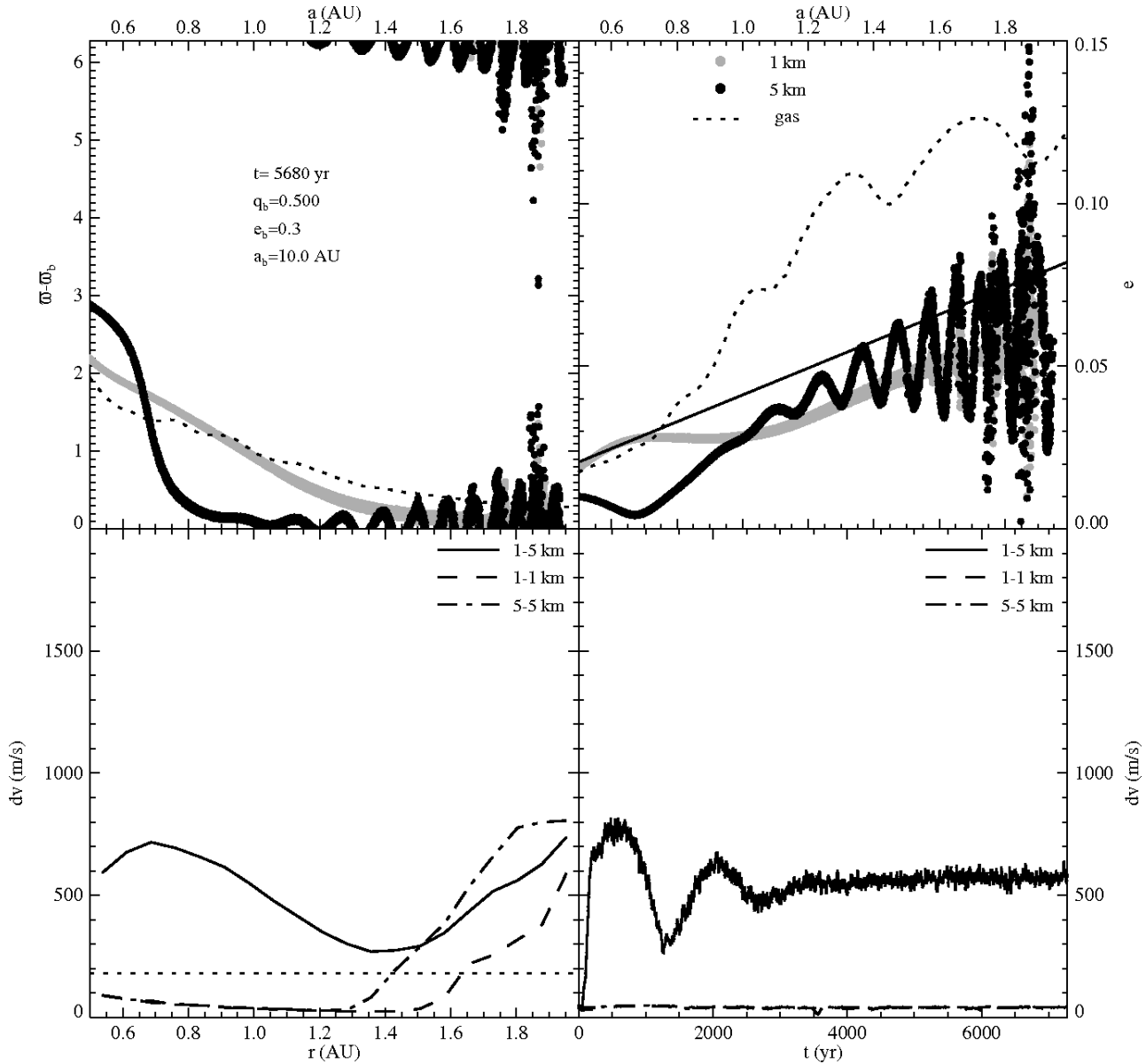}
\caption{Planetesimal evolution in an evolving gas disc (the quiet
state), with a steep radial gas profile in $\Sigma_0 \propto
r^{-7/4}$, for the same tight binary parameters as in
Fig. \ref{figaxigas}. Top left: longitude of periastron distribution
after 5680 yr. Top right: eccentricity distribution after 5680
yr. Bottom left: distribution of encounter velocities for 3 different
impactor-target planetesimal pairs: 2 corresponding to equal sized
impacting objects and the third one to a 1 km impactor hitting a 5 km
target. Also shown is the limiting $\langle dv \rangle$ for accreting encounters (dotted line). 
Bottom right: encounter velocities at 1~AU as a function of time for the same 3 impactor-target pairs.}
\label{figSPmm}
\end{figure*}

\begin{figure*}
\centering
\includegraphics[width=17cm,clip=true]{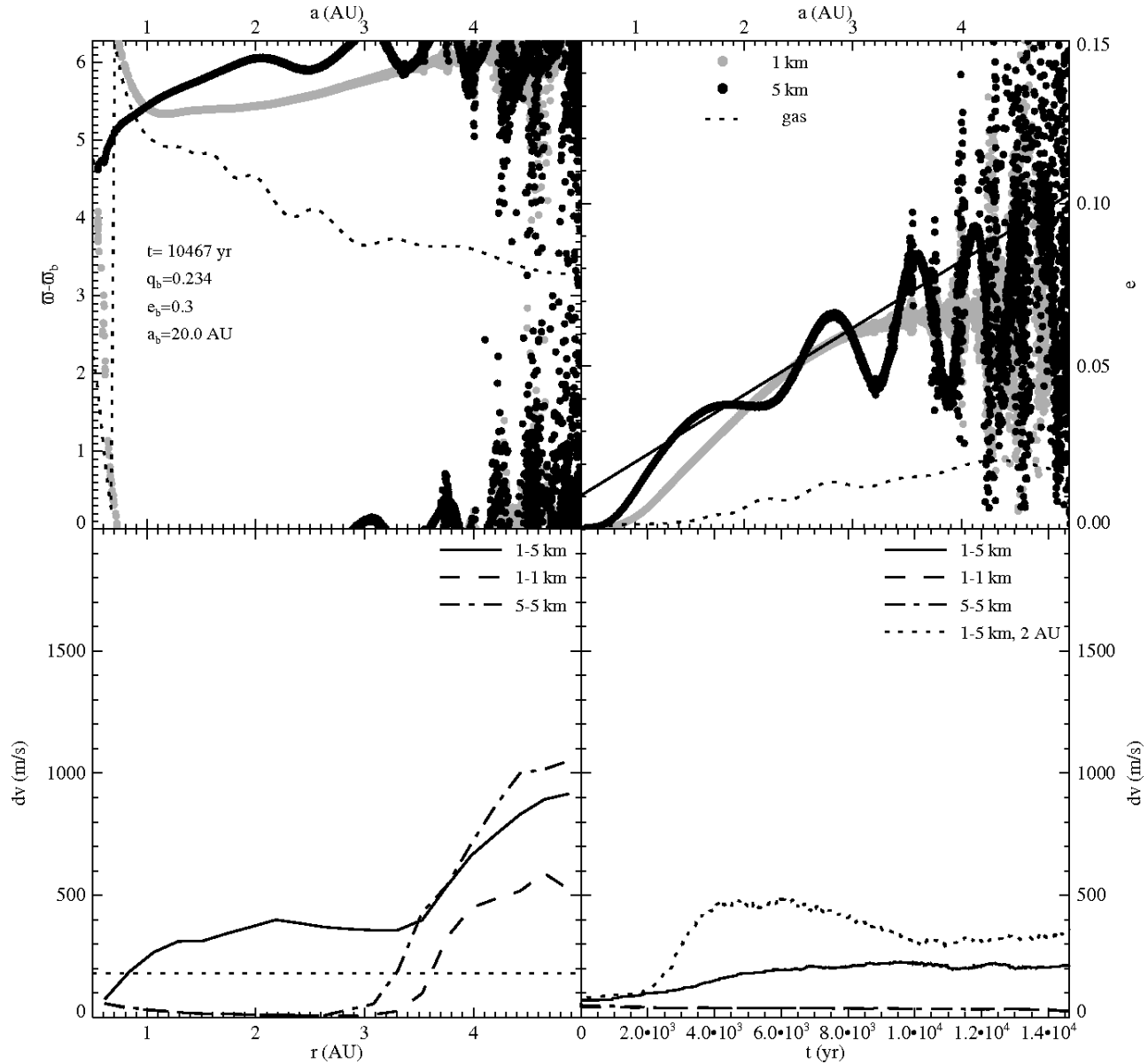}
\caption{Planetesimal evolution in an evolving gas disc (the quiet
state), with a shallower radial gas profile $\Sigma_0 \propto
r^{-1/2}$, and for the wider, $\gamma$~Cephei like binary case. Top
left: longitude of periastron distribution. Top right: eccentricity
distribution. Bottom left: distribution of encounter velocities for 3
different impactor-target planetesimal pairs: 2 corresponding to equal
sized impacting objects and the third one to a 1 km impactor hitting a
5 km target. Also shown is the limiting $\langle dv \rangle$ for accreting encounters (dotted line). Bottom right: encounter velocities at 1~AU as a function of time for the same 3 impactor-target pairs. In addition, we show the
encounter velocities at 2~AU for a 1 km impactor hitting a 5 km target
(dotted line).}
\label{figLPmm}
\end{figure*}

\begin{figure*}
\centering
\includegraphics[width=17cm,clip=true]{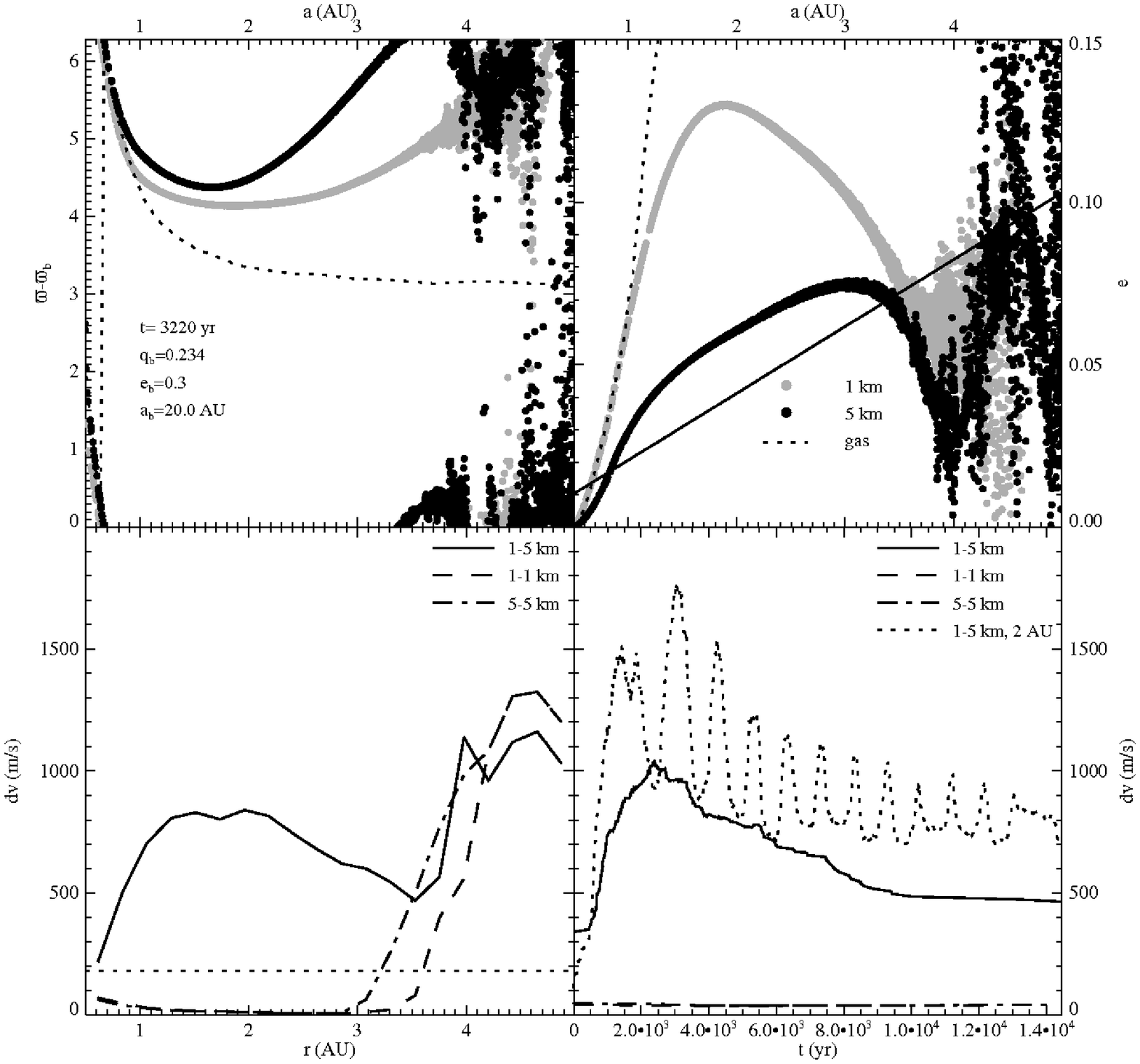}
\caption{Planetesimal evolution in an evolving gas disc (the excited
state), with $\Sigma_0 \propto r^{-1/2}$, for the same binary
parameters as in Fig. \ref{figLPmm}. Top left: longitude of periastron
distribution. Top right: eccentricity distribution. Bottom left:
distribution of encounter velocities for 3 different impactor-target
planetesimal pairs: 2 corresponding to equal sized impacting objects
and the third one to a 1 km impactor hitting a 5 km target. Also shown is the limiting $\langle dv \rangle$ for accreting encounters (dotted line). Bottom
right: encounter velocities at 1~AU as a function of time for the same
3 impactor-target pairs. In addition, we show the encounter velocities
at 2~AU for a 1 km impactor hitting a 5 km target (dotted line).}
\label{figLPsb}
\end{figure*}

The soft superbee flux limiter gives rise to a different disc state,
with a large free eccentricity, reaching values up to $\eg \sim 0.2$.
We will refer to this state as the excited state. This state is
similar to the one discussed in \cite{kleynel07}. Its exact behavior
depends strongly on resolution and boundary conditions, but the
overall picture is the same: high eccentricity, and the disc starts to
precess. This is illustrated in Fig. \ref{figprecess}, where we show
the mass-averaged eccentricity and longitude of periastron for the
disc.  Within 20 orbits, the disc obtains a large eccentricity, after
which the disc starts to precess slowly in a retrograde fashion. The
eccentricity oscillates with the same period as the precession. The
retrograde precession indicates that pressure forces are dominant
\citep[see][]{papa05}. In Fig. \ref{figprecess}, the quiet disc
quickly reaches a steady eccentricity distribution. Note that due to
the twist in the disc (see Fig. \ref{figgasecc}), which is dynamically
unimportant, the average longitude of periastron approaches neither
$0$ nor $\pi$, but somewhere in between.

Note that for both states the disc eccentricity does not equal the
forced eccentricity, denoted by the thick solid line in
Fig. \ref{figgasecc}. For the quiet state, we have checked that the
gas eccentricity approximately follows Eq. \eqref{eqEgas}. The fact
that, despite the density being proportional to $r^{-1/2}$, the gas
eccentricity does not tend to the forced eccentricity is due to the
behavior near the truncation radius. This is basically where the
boundary condition for Eq. \eqref{eqEgas} is determined, which, as
shown in Fig. \ref{figeccana} can easily lead to a strongly reduced
disc eccentricity.  According to the analytical study of
Sect. \ref{secAna}, this means that we can expect large encounter
velocities between planetesimals in both cases.

The reason for the large differences between the runs with different
flux limiters is that the mechanism responsible for generating a large
free eccentricity involves an eccentric resonance near the outer edge
of the disc. Basically, it is the mechanism outlined in
\cite{lubow91}, but for an eccentric companion orbit.  \cite{heems94}
showed that eccentricity excitation through this instability strongly
depends on the way the disc is truncated, which in turn depends
strongly on wave damping. The two different limiters give different
wave damping, hence our results, see Appendix A. Only for small wave
damping does the disc obtain a large free eccentricity. It is expected
that this also depends on the magnitude of the physical viscosity,
since this affects the shape of the disc edge. We provide some
additional comments in Sect. \ref{secDisc}, but leave a detailed
discussion of this problem to a future publication. We now proceed to
analyze the influence of both disc states on planetesimal accretion.

\subsection{Planetesimal evolution}

\subsubsection{Quiet disc case}

We first consider the quiet disc case,
for which the situation should a priori be the closest to the
non-evolving axisymmetric case. 

Fig. \ref{figSPmm} shows the evolution of planetesimal orbital
parameters and encounter velocities for the same tight binary
parameters as in Fig. \ref{figaxigas}. Comparing the top panels of
Figs. \ref{figaxigas} and \ref{figSPmm} we see that, as expected, the
largest differences occur in the inner regions of the disc, typically
within $\sim 0.8$~AU, where gas drag effects are the most
important. The 1-km planetesimals approach the gas eccentricity
towards the inner boundary, as expected. Interestingly, the 5-km
bodies show a large jump in longitude of periastron around $r=0.7$~AU,
accompanied by a drop in eccentricity. From the top left panel of
Fig. \ref{figSPmm} we see that this happens where the longitude of
periastron of the gas disc amounts to $\pi/2$. Around this location,
depending on the value of $Z$, the denominator of Eq. \eqref{eqtanpom}
will approach zero, causing a large jump in $\varpi$. For the 5 km
planetesimals, $Z\approx 3$ at this location, which causes a drop in
eccentricity (see Eq. \eqref{eqe2}). This is not true for the 1 km
planetesimals, and therefore there is a large eccentricity difference
around $r=0.7$~AU.

In terms of encounter velocities, these different behaviours of 1 and
5 km bodies in the innermost regions logically translate into higher
$\Delta v$ than in the axisymmetric case (see the bottom-left panel of
Figs. \ref{figaxigas} and \ref{figSPmm}). At 1~AU, the equilibrium
encounter velocities are approximately a factor of 2 higher than for
the case of a circular gas disc. However, in the outer disc, beyond
$\sim$1~AU, differences with the axisymmetric case are much
smaller. Although the gas eccentricity is higher in these regions,
the gas density is not high enough to significantly affect the
behaviour of km-sized planetesimals. Beyond $r=1.4$~AU, the dynamical
evolution of the planetesimal population becomes indistinguishable
from the circular gas disc case.

We now turn our attention to binary parameters that match those of
$\gamma$~Ceph: $\qb=0.234, \ab = 20$~AU and $e_b=0.3$. From
Fig.~\ref{figgasecc} we see that the gas disc eccentricity is very
small, $\eg<0.02$ almost everywhere. In the top panels of
Fig.~\ref{figLPmm} we show the distribution of longitude of periastron
and eccentricity after $10^{4}$ yr, when the system has reached a
steady state. We find that in the whole $r\geq0.8$AU region, the
equilibrium $\langle \Delta v \rangle$ for collisions between 1 and 5
km objects is $\geq200\,$\mps. This is still high enough to correspond
to eroding impacts for all tested collision outcome prescriptions of
\cite{the06} (see bottom-left panel of Fig.\ref{figLPmm}). Direct
comparison with Fig. \ref{figaxigas} is here difficult, because the
binary parameters are different. We thus performed an additional
axisymmetric gas disc test simulation which showed that the encounter
velocities are the same, within $10 \%$, as for the present quiet
state run. This is an indication that the spiral waves, that do extend
all the way in, are indeed of minor importance regarding impact
velocities. The circular gas disc case is then a relatively good
approximation.

\subsubsection{Excited disc case}

In Fig. \ref{figLPsb} we show the results for the excited disc state,
for the $\gamma$~Cephei like binary. It is immediately clear that,
even for this wide binary case, the planetesimals react rather
violently to the large eccentricity of the gas disc. Inside 1~AU, the
1~km planetesimals are dragged along with the gas and end up on highly
eccentric orbits matching that of the gas. The 5~km planetesimals
follow the same trend, but are more loosely coupled to the gas
streamlines and their eccentricities never match that of the gas. For
the periastra, however, we observe a quasi-perfect alignment with that
of the gas for both planetesimal sizes. Note that this rather abruptly
happens where $\varpi_\mathrm{g}\approx 3\pi/2$, which makes
$|\tan(\varpi_\mathrm{g})|$ very large. From Eq. \eqref{eqtanpomg} we
see that in this case, for which we indeed have $\eg \gg \ef$, we
expect all planetesimals to align with the gas, independently of $Z$.
Outside 1~AU, 1 km planetesimals start to decouple from the gas, while
the bigger bodies do so already at $\sim 0.7$~AU. In the whole $1\leq
r \leq 3\,$AU region, differences between the equilibrium
eccentricities for both sizes are very large, culminating at $\sim
2\,$AU, where $e_\mathrm{1km}\approx 2\,e_\mathrm{5km}$.

\begin{figure}
\resizebox{\hsize}{!}{\includegraphics[clip=true]{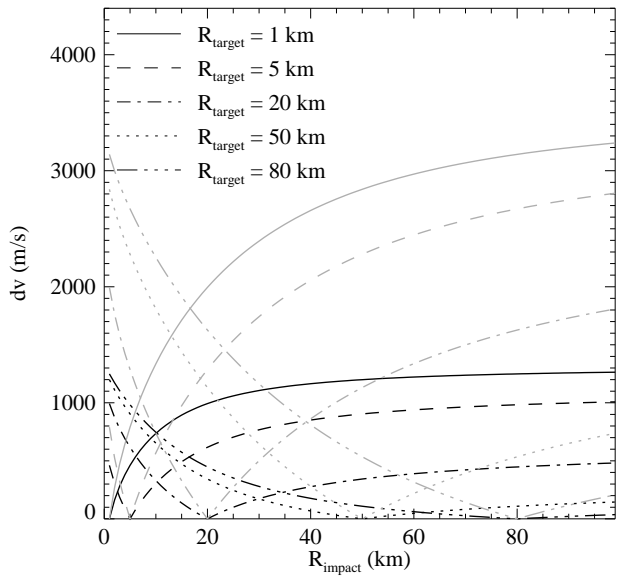}}
\caption{Encounter velocities, calculated using Eq. \eqref{eqColl} for
the mean eccentricity of the excited gas disc state, for binary
parameters $\qb=0.234$, $\ab=20$~AU and $e_b=0.3$. Black lines were
calculated using $\Eg=0.05$, appropriate for 1~AU (see
Fig. \ref{figgasecc}), grey lines using $\Eg=0.15$, appropriate for 2
AU.}
\label{figcollecc}
\end{figure}

The differences in $E$ between the two particle sizes in the
planet-forming region around 1~AU lead to very high encounter
velocities, exceeding $500\,$\mps\ almost everywhere (bottom panel of
Fig. \ref{figLPsb}). Note that due to the precession of the disc,
$E_\mathrm{g}$ is time dependent, as appears clearly from the
oscillations in Fig. \ref{figprecess}. However, this precession
timescale is shorter than the time it takes for the planetesimals to
settle into their equilibrium distribution. Therefore, the planetesimals
will feel an average gas eccentricity and eventually settle into a
steady eccentricity and $\langle \Delta v \rangle$
distribution. Also, the disc at 1~AU does not really participate in
the precession, possibly due to the fact that it is close to the inner
boundary (see also below). This is also apparent from
Fig. \ref{figgasecc}. For comparison, we also show the encounter
velocities evolution at 2~AU in the bottom right panel of
Fig. \ref{figLPsb}. We see that, at this larger distance, the
encounter velocities oscillate, with a period which is that of the gas
disc precession (see Fig. \ref{figprecess}), but after approximately
$10^{4}$yrs a meaningful average can be defined. We note that, at a
given location in the disc, encounter velocities are close to those
predicted by Eq. \eqref{eqColl}, when plugging in the average value of
$E_\mathrm{g}$ at this location (see also Fig. \ref{figcollecc}). This
again shows that it is the eccentricity of the gas disc that governs
the impact velocities.

The bottom left panel of Fig. \ref{figLPsb} shows that encounter
velocities in the inner regions, which are due to (gas drag induced)
differential orbital phasing, come close to those in the outer disc,
which are due to pure gravitational orbital crossing. For 1km-5km
pairs, impact velocities are on average 5 times higher than those in
the circular disc run. They are thus far above the upper limit for
accreting impacts for the whole simulated region $0.6<r<6\,$AU.

We have also run the soft superbee flux limiter for the tight binary
case. Differences with the circular case are slightly larger than
for the quiet state, but the gas disc is not able to obtain a large
free eccentricity in this case, for two reasons. First, numerical and
analytical experiments suggest that the maximum eccentricity (free or
forced) that the gas disc can reach is largely determined by the
boundaries (see also Fig. \ref{figeccana}). The quiet state of the
tight binary comes close to obtaining this maximum of $\eg \sim 0.3$,
and therefore the excited state is not very much different in this
case. Second, because the gas disc is truncated closer to the primary,
due to the stronger perturbations from the tight binary, the mechanism
for generating a large free eccentricity is reduced in strength
compared to the wide binary case.  Therefore, encounter velocities do
not vary more than $50\, \%$ between the quiet and excited state for
the tight binary, meaning that velocities are here a factor of
$\sim 3$ higher than in the circular gas disc case.

As already mentioned, numerical experiments have shown that the
position of the inner boundary determines up to what radius the gas
disc obtains its large eccentricity. For example, in
Fig. \ref{figgasecc}, the eccentricity drop inward of 2~AU is directly
connected to the location of the inner boundary at $0.25\,$
AU. Exploring this parameter is outside the scope of the present
paper, since moving the inner boundary further inwards is
computationally very expensive. However, it is important to note that
it would only act to further \emph{increase} the encounter velocities.
More work is necessary to study the behavior of the excited disc
state close to the inner edge of the disc, numerical or
physical. Here, we focus on the influence of a gas disc with a large
free eccentricity on planetesimal collisions only, and we have shown
that it is disastrous for planetesimal growth.

\section{Discussion}
\label{secDisc}

\subsection{The impact of gas disc eccentricity on encounter velocities}

Our results indicate that planetesimal encounter velocities, which
result from differential orbital phasing due to gas drag, critically
depend on the eccentricity of the gas disc. Although there is a
theoretical possibility for differential phasing to be avoided, when gas streamlines follow the forced
dynamical orbits, we have shown that this situation does not occur in practice. First of all,
Eq. \eqref{eqEgas} shows that we can only have $\Eg=\ef$ when the gas
density follows $\Sigma \propto r^{-1/2}$. Even if this happens to be
the case, there is a strong dependence on what happens at the inner
and outer boundaries of the disc (see Fig. \ref{figeccana}), in
combination with the disc thickness. More importantly, this $\Eg=\ef$
condition was never encountered in our numerical hydrodynamical
simulations. Since we could not explore all free parameters, we cannot
rule out the possibility of obtaining $\Eg=\ef$ very locally when
tuning in the right disc parameters, but this would have to be a
fortuitous coincidence. Even more, if the disc reaches a large free
eccentricity, then it will necessarily precess, so that $\Eg=\ef$
would not be a steady state.

Not only does this $\Eg=\ef$ state never occur, but for all 4 cases
here explored (2 flux limiters + 2 binaries configurations), we always
obtain, at all radial distances, $\langle \Delta v \rangle$
equilibrium values \emph{higher} than for a static circular gas disc
case. For the quiet state, differences with the static case remain
relatively limited: velocity distributions are almost identical for a
binary with $20\,$AU separation and are enhanced by slightly less than
a factor of 2 for a tight $\ab=10$\,AU companion. For the excited
state, however, differences with the static disc case can become
large. They already exceed a factor of 2 for the wide (20\,AU) binary
case and reach a factor of 3 for the tight binary.  Interestingly,
even for the case where $\eg$ turns out be relatively close to $\ef$,
i.e., in the inner disc in the quiet state/tight binary run (see
Fig. \ref{figSPmm}), we still observe an important differential
phasing between particles of different sizes and thus high $\langle
\Delta v \rangle$ values. This is entirely due to the fact that the
gas disc is not aligned with the binary in that region. This is a
further indication that the parameter phase space around the exact
$\Eg=\ef$ condition where differential phasing is weaker than for a
circular disc is very limited.

The computational cost of the presented numerical simulations inhibits
an extended parameter study similar to the one in \citet{the06} for
the static circular gas disc case. As already mentioned, we had to
restrict ourselves to two representative binary configurations as well
as to two planetesimal sizes, 1 and 5\,km.  However, we have checked
(see Fig.\ref{figcollecc}) that for all numerically explored cases,
Eq. \eqref{eqColl} gives good estimates of equilibrium encounter
velocities once given the value for the mean gas disc eccentricity
$\Eg$ (which is obtained from Fig. \ref{figgasecc} and an equivalent
plot for the quiet state). As a first approximation, this equation can
thus be used as a tool to estimate $\langle \Delta v \rangle$ between
planetesimal of any sizes $R_1$ and $R_2$, for any given binary
orbital configuration. The only unknown input parameter is $\Eg$ at a
given location in the disc, but this value can be retrieved from
independent pure hydro-simulations, which are less time consuming than
full gas+planetesimal runs.
 
For the present binary configurations, we use Eq. \eqref{eqColl} to
estimate impact velocities for planetesimals bigger than the ones
explored in the runs (see Fig.\ref{figcollecc} for the excited
case). This could be of interest for two reasons:
\begin{enumerate}
\item Sizes in the
``initial'' planetesimal population, however fuzzy this concept might
be, are very poorly constrained by planet-formation scenarios.
\item The
steady state with equilibrium $\langle \Delta v \rangle_{R1,R2}$ is
not reached instantly and might take a few $10^{3}\,$yrs to settle
(see Figs. \ref{figSPmm} and \ref{figLPsb}), which might leave time
for some planetesimals to start to grow and reach bigger sizes.
\end{enumerate}
\citep[for more detailed discussions on these 2 issues, see][]{the06}.
From Fig. \ref{figcollecc}, we see that, for the $\gamma$~Cephei-like
binary, impact velocities quickly rise to very high values for larger
bodies. Even though exact collision outcomes for such large objects
remain poorly constrained, such extremely high velocities are very
likely to lead to mass erosion or shattering of the impactors. As an
example, for the same binary configuration (except a slighly higher
$\qb$ value) and for collisions between 20 and 50 km bodies,
\citet{the06} find $\langle \Delta v \rangle \sim 300\,$\mps\ with a
circular gas disc, a value for which no clear conclusions in terms of
accreting vs. erosive impacts can be drawn (see Fig.\,9 of that
paper). For the present excited case, however, values larger than
$10^{3}\,$\mps\ are reached, which are without doubt high enough to
lead to net mass loss after the impact.

\subsection{The origin of gas disc eccentricity}

Additional test simulations with a circular binary orbit showed
similar behavior as for the eccentric binary case. Depending on the
amount of wave damping, the disc can quickly become eccentric. We have
checked that in this case it is the 3:1 eccentric Lindblad resonance
\citep{lubow91} that drives the eccentricity. Removing the $m=3$
component from the companion potential resulted in almost circular discs
for all flux limiters. This indicates that a similar mechanism
operates in the case of an eccentric binary. When the disc does become
eccentric, the circular binary case behaves in a similar fashion as
the eccentric binary case. This is because $|E_\mathrm{g}| \gg
e_\mathrm{f}$: the eccentricity evolution of the planetesimals is
dominated by the free eccentricity of the gas disc. Whether the
3:1 resonance can induce a large eccentricity in the gas disc strongly
depends on the truncation of the disc. It is not expected to happen
for equal-mass binaries \citep{lubow91}, which was the case studied by
\cite{pawel07} for a non-evolving gas disc. More work is necessary to
find out at which mass ratio this instability sets in and what
values of $\eg$ can be reached.

We have not found evidence for a viscous overstability operating in
the disc. However, there exists a complex interplay between viscosity,
tidal effects and eccentricity excitation. If the disc is truncated
inside the main eccentricity-generating resonance, the disc will not
obtain a large free eccentricity. However, a strong viscosity may
spread the outer edge of the disc into the resonance, and this
complex problem clearly deserves more study. Our formalism, in
particular Eq. \eqref{eqColl}, allows for a direct link between gas
disc eccentricity and planetesimal encounter velocities, at least in
the region where orbital crossing can be neglected. This may ease
the computational cost of future studies, as simulations with gas only
are required to investigate planetesimal encounters.

An important parameter governing gas friction is the ambient gas
density. In our simulations, we always rescale the density at 1~AU to
$\rho_\mathrm{g}=1.4 \cdot 10^{-9}$ g $\mathrm{cm}^{-3}$, in order to
compare with previous work \citep{the06}, the actual density may be
very different. Even for single stars the disc mass is relatively
poorly constrained, and the effect of the binary further complicates
things. As the gas density only appears in the drag force parameter
$K$, which also contains the planetesimal size, our results can always
be scaled to different gas densities by considering different
planetesimal sizes.

The effects of self-gravity are usually neglected in calculations of
protoplanetary discs, because unless the disc is very massive
($M_\mathrm{d}\sim 0.1\,$$\ms$) its dynamical influence is small. The
parameter measuring the importance of self-gravity is the Toomre $Q$
parameter \citep{toomre}:
\begin{equation}
Q=\frac{c_\mathrm{s}\kappa}{\pi G\Sigma},
\end{equation}
where $\kappa$ is the epicyclic frequency. For a locally isothermal
Keplerian disc with a constant aspect ratio we have:
\begin{equation}
Q=\frac{h}{\mu_\mathrm{D}},
\end{equation}
where
\begin{equation}
\mu_\mathrm{D}=\frac{\pi r^2\Sigma}{\ms}
\end{equation}
is a measure of the disc mass. Taking a disc mass of the order of one
Jupiter mass we find $Q\approx 50$, which indicates that self-gravity
is not important. However, Toomre's stability criterion is based on
short wavelengths. For global modes, self-gravity can dominate over
pressure much more easily \citep{papa02}. Basically, the parameter
measuring the importance of self-gravity becomes $hQ$, which is of
order unity for our model disc. Therefore, future models should
include self-gravity of the gas.

Within our analytical framework, taking into account
gravitational forces due to the gas amounts to adapting the forcing
potentials appearing in Eq. \eqref{eqEnodrag} to include perturbations
due to the gas. These will involve integrals of the gas surface
density over the disc, where for $\Phi'_2$ the integrand will depend
on $\Eg$. For large enough disc mass and disc eccentricity, the
forcing due to the eccentric gas disc will dominate over the binary
forcing. We may expect this to be the case in the excited
state. Although a detailed analysis of this case is beyond the scope
of this paper, we comment that we do not expect this case to differ
qualitatively from the case studied here. Again, there will be an
equilibrium eccentricity distribution for the gas, which will be
harder to compute because now an integro-differential equation has to
be solved, and an equilibrium eccentricity distribution for the
planetesimals. In general, these will be different due to the spatial
derivatives appearing in the gas eccentricity equation, which will
give rise to differential orbital phasing. Again, there may be a
special case for which $\Eg=E_0$ for all planetesimal sizes, but it is
not clear whether this state is reached more easily than for the cases
studied here. This clearly deserves more study.
   

\section{Summary and conclusion}
\label{secCon}
We present the first simulations of planetesimal dynamics in binary
systems including full gas disc dynamics. As in past studies with
static axisymmetric gas discs, we confirm the crucial role of
differential orbital phasing due to gas drag in inducing large
encounter velocities between bodies of different sizes. Interestingly,
while there is a theoretical possibility for differential phasing to
be less pronounced than in the axisymmetric case (if $E_g \sim e_f$),
our numerical exploration shows that this case never occurs in
practice: we always find stronger differential phasing than for a
simplified static circular gas disc.

The level of the phasing is connected to the eccentricity reached by
the gas disc. Depending on the amount of wave damping, the disc either
enters a nearly steady state, for which encounter velocities are
within a factor of 2 difference from the case of a circular gas disc,
or the disc enters a highly eccentric state with strong precession,
for which the encounter velocities can go up by almost an order of
magnitude in the most extreme cases. It is important to point out
that, for both cases, the global qualitative effect is always to
\emph{increase} $\langle \Delta v \rangle_{(R1,R2)}$ for $R_1 \neq
R_2$.

Taking into account the gas disc's evolution thus leads to a dynamical
environment which is less favourable to planetesimal accretion. In
this respect, despite of the fact that the high CPU cost of these
simulations prevented us from performing a thourough exploration of
all binary parameters, it is likely that the ($\ab$,$\eb$) space for
which planetesimal accretion is possible is more reduced than was
found in previous studies with an axisymmetric disc
\citep[e.g.][]{the06}.

\bibliography{paard.bib}


\appendix

\section{Flux limiters and eccentric discs}
\label{secappendix}

In view of the large influence of the choice of flux limiter on the
resulting disc structure in a binary system, it is appropriate to
review their use in numerical hydrodynamics. Our discussion closely
follows \cite{leveque}. For most problems the choice of flux limiter
does not play a major role and thus it is not common practice to check
the effect of the choice of flux limiter. Our results show that it
would be wise to always do so, especially in highly non-linear
problems.

\subsection{Linear advection equation}
We keep the discussion simple by considering the linear advection
equation, in stead of the non-linear system of Euler equations:
\begin{equation}
\label{eqLinAdv}
\frac{\partial u}{\partial t}+a\frac{\partial u}{\partial x}=0,
\end{equation}
where $u$ is the conserved quantity (eg. mass, momentum, energy) and
$a$ is the advection velocity, which we take to be a positive
constant. We will denote the numerical solution, discretized on an
equidistant grid $\{x_i\}$, at time step $n$ as $\{U_i^n\}$. An
important difficulty arises in numerically solving this type of
equation, because it allows for weak, or discontinuous, solutions. The
infinite gradients at discontinuities are problematic for numerical
difference schemes, leading to strong oscillations in the numerical
solution. In fact, a theorem due to \cite{godunov} states that a
linear method that preserves monoticity (i.e. does not lead to
oscillations) can be at most first-order accurate. Therefore, while in
regions of smooth flow it is desirable to have a scheme that is
second-order, in the presence of discontinuities or shocks it is
necessary to switch to a first-order scheme. This is where the flux
limiter enters the stage.

Suppose we have a method for determining the numerical flux $F$ across
a cell boundary. Moreover, we should have a second-order flux, $F_2$,
and a first-order flux, $F_1$. In regions of smooth flow, we want to
use $F_2$ for higher accuracy, while near a discontinuity we want to
use $F_1$ to avoid spurious oscillations. To achieve this, we write
the total flux as:
 \begin{equation}
 \label{eqNumFlux}
 F=F_1+\Phi(\theta)\left(F_2-F_1\right),
 \end{equation}
 where $\Phi(\theta)$ is the flux limiter (not to be confused with the
 potential $\Phi$ in the main text). In regions of smooth flow, $\Phi$
 should be close to 1, while near discontinuities we want $\Phi=0$ to
 switch to the first-order solution. Given a measure of the smoothness
 $\theta(U)$, it is possible to obtain bounds on the functional form
 of $\Phi(\theta)$ to prevent spurious oscillations near shocks
 \citep{sweby}. Within these bounds, one is free to choose the flux
 limiter, and for all choices one obtains a method that is
 second-order in regions of smooth flow, and does not introduce
 oscillations near shocks. For Eq. \eqref{eqLinAdv}, a natural choice
 for $\theta$ is the ratio of two consecutive gradients:
\begin{equation}
\theta_i=\frac{U_i-U_{i-1}}{U_{i+1}-U_i}.
\end{equation}
Note, however, that this measure of smoothness breaks down near
extreme points of $U$.
 
A different, and more graphical, interpretation of $\Phi$ is to regard
it as a slope limiter. To evolve $U$ for a time $\Delta t$, we can
regard $U$ as a piecewise constant function, shift it by an amount of
$a\Delta t$, and finally integrate to obtain new cell averages. This
leads to the upwind method (note that we assume $a>0$):
\begin{equation}
U_i^{n+1} =U_j^n-\epsilon \left(U_i^n-U_{i-1}^n\right),
\end{equation}
where $\epsilon=a\Delta t/\Delta x$ is the Courant number. For
numerical stability, we should have $\epsilon<1$. The upwind scheme is
a first-order method, and does not produce oscillations near
discontinuities.

\begin{figure}
\resizebox{\hsize}{!}{\includegraphics[clip=true]{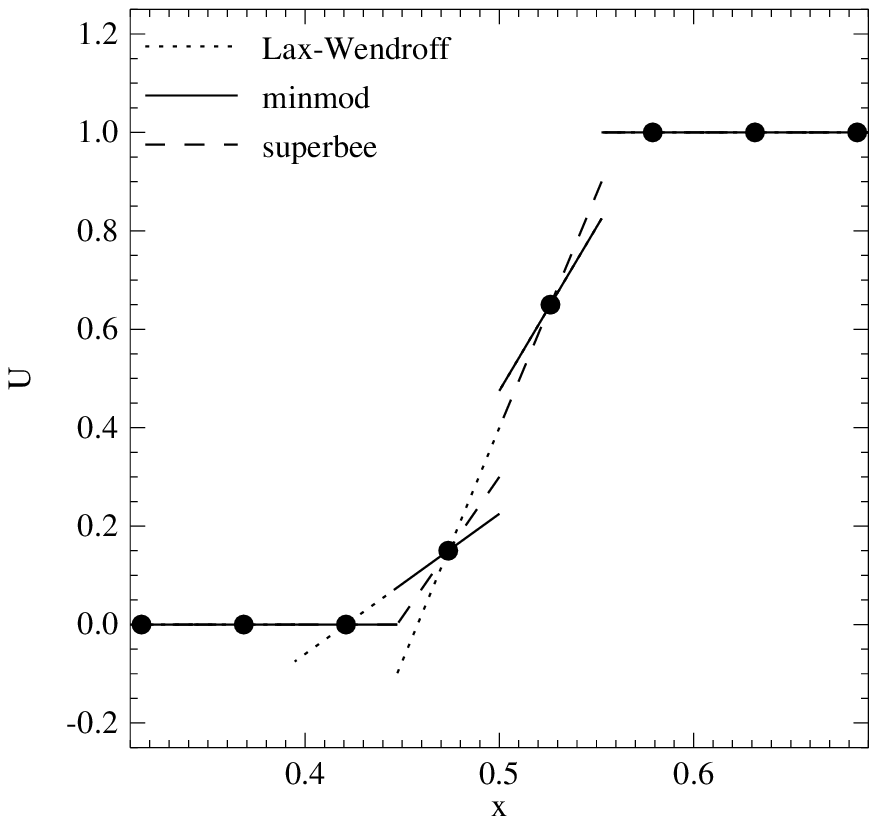}}
\caption{Numerical solution $U$ of Eq. \eqref{eqLinAdv} at time $t_n$,
and $\Delta x=0.05$ (dots), together with the resulting interpolation
for three different slope limiters.}
\label{figslopelim}
\end{figure}

We can improve on the upwind scheme by releasing the assumption of $U$
to be piecewise constant, and in stead allow $U$ to vary linearly
within one grid cell with slope $\sigma$. Applying the same technique
(shifting and integrating this new $U$) leads to a different update
for $U$:
\begin{equation}
U_i^{n+1} =U_i^n-\epsilon \left(U_i^n-U_{i-1}^n\right)-\frac{1}{2}\Delta x\epsilon(1-\epsilon)\left(\sigma_i^n-\sigma_{i-1}^n\right).
\label{eqSlopeLim}
\end{equation}
For $a>0$, if we make the natural choice
$\sigma_i=(U_{i+1}-U_i)/\Delta x$, Eq. \eqref{eqSlopeLim} reduces to
the Lax-Wendroff method, which is second-order but leads to
oscillations near shocks. If we in stead use
\begin{equation}
\sigma_i=\frac{U_{i+1}-U_i}{\Delta x}\Phi(\theta),
\end{equation}
we end up with a numerical flux of the form of Eq. \eqref{eqNumFlux},
in which $\Phi$ can now be interpreted as a slope limiter. In
Fig. \ref{figslopelim} we show the difference between three different
slope limiters near a sharp gradient. The Lax-Wendroff slope (dotted
lines) is obtained by setting $\Phi=1$. The minmod and superbee
limiters are examples of a general class of limiters of the form
\begin{equation}
\Phi(\theta)=\max(0,\min(1,p\theta),\min(\theta,p)),
\end{equation}
where for $p=1$ the limiter is called minmod, and for $p=2$ we have
the superbee limiter. In the main text, we have used $p=1.5$, which we
call the soft superbee limiter. For $1\leq p \leq 2$, it can be shown
that this limiter prevents oscillations in the numerical solution near
discontinuities. Away from the sharp gradient, all limiters produce the same result. 

From Fig. \ref{figslopelim}, it is clear that the Lax-Wendroff slope
will introduce oscillations near $x=0.45$. The other two limiters will
keep the solution monotonic, but with different slopes. In general,
the superbee limiter will give rise to the steepest slopes possible
without introducing oscillations in the solution. This way, numerical
smearing of shocks is minimal, but at the same time, shallow gradients
may be steepened artificially. In practice, it is safer to use a
``softer'' limiter, where one would expect that minmod will give results
that come closest to those obtained with non-shock-capturing schemes,
which need explicit artificial viscosity to stabilize shocks.

\subsection{Application to eccentric discs}

We expect the influence of the flux limiter to be largest near
discontinuities, since all limiters give $\Phi=1$ in regions of smooth
flow. The problem at hand introduces strong non-linearity in the disc
response. The spiral waves induced by the binary companion are in fact
spiral shocks, and their strength is directly related to the growth of
eccentricity in the disc \citep{lubow91}. Moreover, their dissipation
is related to the location at which the disc is truncated
\citep{linpap86}, which also affects the strength of the instability
\citep{heems94}. Finally, when $\eg>h$, supersonic radial velocities
occur. Since usually $\eg\rightarrow 0$ near the boundaries of the
computational domain, this will introduce shocks in the problem as
well.

It is clear, therefore, that shock dissipation plays a crucial role in
the growth as well as in the damping of eccentricity in the gas
disc. A diffusive flux limiter will reduce the effectiveness of
eccentricity excitation, and will damp the obtained eccentricity more
readily. Our results in the main text indicate that this leads to the
mechanism involving the 3:1 Lindblad resonance being unable to induce
eccentricity in the disc when the minmod limiter is used. This strong
influence of the flux limiter is remarkable, and indicates that
strongly non-linear effects play a large role in this problem. Also,
it is interesting that non-shock-capturing methods seem to reproduce
the results obtained with the soft superbee flux limiter rather than
those obtained with the minmod limiter \citep{kleynel07}. The limiter
used by \citet{kleynel07} is the so-called van Leer limiter, which
actually has a functional behaviour quite close to our soft
superbee. However, the basic numerical algorithm for solving the
hydrodynamic equations differ considerably between our method and
those used in \citet{kleynel07}, so it is somewhat too simplified to
compare them at the level of limiter functions. This clearly deserves
more study. In the main text, we have studied both cases, using the
minmod limiter to study a case for which the 3:1 Lindblad resonance
does not operate.


\label{lastpage}

\end{document}